\documentclass[a4paper,11pt]{report}
\usepackage{amssymb}
\usepackage{makeidx}
\usepackage{indentfirst}
\usepackage{amssymb,amsfonts,graphicx,amssymb,mathrsfs,amsmath,bm,dcolumn}
 \makeindex

\begin{document}

\title{\textbf{Dark Energy, Hyperbolic Cosecant Cardassian and\\
Virial Collapse for Power-style Cardassian}}
\date{}
\author{Wen-Jie Tian\footnote{tianwj1@gmail.com}\\
College of Physics and Information Technology,\\ Shaanxi Normal
University, Xi'an, 710062, P.R.China} \maketitle

\tableofcontents
\newpage
\begin{center}
\textbf{\Large{Abstract}}\\
\end{center}

The problem of dark energy(DE) is the greatest challenge for modern
physics, and therefore this thesis is dedicated to the modeling
establishment of DE. In the first chapter, qualitative and
descriptive methods are employed to draw a rough map for the cosmos
in its finely analyzed era. Then the dynamical equations of
cosmology based on the general relativity are introduced, together
with the measuring methods and parameters. The second chapter
reviews the results of cosmic microwave background radiation and
supernova observations, which conclude the flat topology and
accelerating expansion of cosmos, and give rise to existence of DE
and the associated dynamical mechanism problem. Taking WMAP data as
an example, we introduced the practical parameters. In the end we
classify the existing DE models into three categories for further
discussion. The first two chapters make up the basis for the whole
thesis.\\

Chapter 3 reviews the fundamental structures and research progress
of two scalar fields, Quintessence and Phantom, which correspond to
the case of EOS beyond -1 and below -1 respectively. Chapter 4
reviews the fundamental structures and research progress of another
two scalar fields, Phantom and Quintom with non-canonical
lagrangian, both of which have EOS crossing -1. But Phantom field is
real, while Quintom complex. Chapter 5 reviews two non-scalar-field
DE formalism of K-essence and Chaplygin gas, which are both rooted
in modern quantum field theory. Totally six kinds of DE is analyzed,
all of which have positive energy density, negative pressure and
could drive the lately accelerating expansion. And they are compared
and identified
in details.\\

New work gathers in Chapter 6 and Chapter 7. Firstly we make a
comprehensive analysis of the existing work, pay particular
attention to Gondolo and Freese's idea of treating Cardassian energy
term as relativistic perfect fluid(GF fluid), point out that a
potential Cardassian term should meet three conditions, and review
three existing Cardassian terms (power style, polytropic style and
its modification, exponential style and its modification), and
eventually put forward the newly found hyperbolic cosecant
Cardassian.\\

Then the Cardassian dynamical equations are introduced generally and
logically under GF fluid scenario, together with the flowing process
of constructing phase space and differential dynamical systems from
Friedmann equation. Hyperbolic cosecant Cardassian term is employed
for concrete computing. The analysis proceeds in two cases, namely a
unified description of matter and radiation energy density (case 1)
and a separate description of matter and radiation terms (case 2).
Formalism of case 2 is more exact at the expense of more
complicatedness, and due to the mathematical symmetry of matter term
and radiation term in hyperbolic cosecant function, the differential
dynamical equations are considerably simplified. Phase space and
dynamical systems for both cases are achieved. When we calculate the
critical points for case 2, amazingly interesting behaviors of
self-consistency and auto-normalization are exhibited, which is a
strong support for the new model,
along with a forever positive sound speed.\\

The process of virial collapse in Cardassian cosmos is analyzed.
Power-style Cardassian term is employed for its simplicity.
Calculation declares that virial collapse of matter alone is
forbidden. Yet Cardassian has excellent ability for virial collapse,
after the virial collapse ending up with a stable sphere, the ratio
of the ultimate radius to the original radius depends on the
adjustable parameters in Cardassian term. And, the mixture of GF
fluid and matter could conduct virial collapse, the ratio of the
ultimate radius to the original radius depends on the adjustable
parameters in Cardassian term, too. No singularity is generated.\\

The creative work in this thesis incudes the introduction of
hyperbolic cosecant Cardassian, which is the fourth ever available
Cardassian term ever found, after the introduction of exponential
style in 2005. This work, along with the calculating of power
Cardassian virialization, enriches the research of Cardassian
cosmology.\\

\chapter{Introductions }
\section{Fundamentals of Cosmology}
Cosmology, which ranks one of the most encouraging realms in natural
science, is the scientific branch researching into the cosmos
structure, origin and evolution on the base of astronomic data and
physics principles. Modern cosmology is basically supported by three
events, namely the cosmic principle, the expansion of the universe
together with Hubble Law, and the Cosmic Microwave Background
Radiation.
\subsection{the Cosmic Principle}
Shortly afterwards the establishment of General Relativity, A.
Einstein computed the global behaviors of the cosmos. To stay in
harmony with Mach Principle and to simplify the calculating process,
he made a basic assumption, which is nowadays honored as the
\emph{Cosmos Principle}. It declares that the universe is
homogeneous and isotropic on large scales. Under modern
considerations, the so-called \emph{large scale} means covering a
distance of $10^8$ light year or more, which belongs to the cosmic
scale. \emph{Homogeneous} means different points identify with each
other and could not be tell apart via physical experiments, while
\emph{isotropic} refers to the fact that , when taking observations
into arbitrary directions at any coordinates origin, the scenarios
are the same. A most significant application of this principle is to
simplify the form of the space-time metric quantitatively. At the
present days, the Cosmic Principle has already been established
firmly on the base of observation data. As a matter of a typical
sample, the results of the Willkinson Microwave Anisotropic Probe
(WMAP) indicate that the universe is in high homogeneity and
isotropy, only with tiny fluctuations at the magnitude of
$10^{-5}\sim 10^{-4}$, which is in fact essentially required for the
generation of galaxies according to the standard model of cosmology.
\subsection{Cosmos Expansion \& Hubble's Law}
In 1929, Hubble was engaged in comparing the light spectrum of far
distant galaxies. To his surprise, he discovered a systematic
redshift with the galaxies beyond our milky-way. As to
\emph{redshift}, it means the whole spectrum translates to the red
wave interval. According to Doppler Effect, all galaxies are leaving
us, and the spatial scale of the universe is undoubtedly expanding.
After some algebra with the redshift amount and the distance to the
galaxies, Hubble found the routing that the farther they are, the
quicker they leaves, at a speed $v$ proportional to the  very
distance $r$:
\begin{equation}
v=H_0r
\end{equation}
where the coefficient $H_0$ is honored Hubble Constant, which is
simply the present-day value of the more generalized dynamical
Hubble parameter.Currently, the results mainly lie in the interval
$H_0\approx 60\sim70 Km\cdot s^{-1}\cdot Mpc^{-1}$.
\subsection{Cosmic Microwave Background Radiation}
The background space of galaxies is a radiation field that is highly
homogeneous and isotropic, whose power spectrum curve agrees with
that of the $2.7K$ blackbody perfectly. The so-called Cosmic
Microwave Background Radiation (CMBR) is a key tool to understand
the structure, origin and evolutionary history of the universe. It
was researching into CMBR that lead to the era of finely
quantitative analysis of cosmology.\\

CMBR could translate quite a lot of information on the young cosmos.
For instance, the perfect homogeneity and isotropy suggest that the
cosmos is also homogeneous and isotropic at its youth era, and the
scattering surface is spherical, on which the gas are of identical
temperature. The background photons are in thermal equilibrium
before decoupling, thus the spectra, which describ how the radiation
strength depending on the power at arbitrary direction, are all
blackbody-like. The temperature fluctuations at present-day
represent the medium density fluctuations at early time, which
magnify via gravitation and form the clustering structure, or the
inhomogeneity on the small scale. CMBR plays a pretty important role
in the standard model of modern cosmology.
\subsection{ Global Scenario}
The standard model of modern cosmology comprises two parts: One is
the heat Big Bang theory, which was put forward by G. Gamow at
around 1948, the other is the inflation theory, which is born in
early 1980s when Guth,Linde, Albrecht and Steinhart employed this
mechanism to beat the difficulties that Big Bang ran into at
extremely early era.\\

The heat big bang theory is based on the observational evidences of
Hubble Expansion and the Helium-4 amounts, and the theoretical
formalism of General Relativity and solar heat nuclear reaction. It
succeeds in explaining the genesis of light chemical elements,
Hubble redshift and forecast the $5K$ CMBR (although the measured
value being 2.7K, this prediction is still a greatest success of
mankind intelligence). The big bang formalism accounts for the
process from $10^{-2}$ second (T $\sim 10^{11}K$) to current time
($\sim 10^{17}s$,2.7K), but falls in trouble with the so-called
singularity problem, the horizon problem, the asymmetry of baryonic
numbers, the flat state problem and origin of galaxies, all of which
are concerned with the extremely early era of the cosmos.\\

The inflation theory is based on general relativity, as well as the
standard model for high energy particle physics established at that
time. It aims at the time scale from $10^{-2}$ second dating back to
$10^{-43}$ second (Plank time), and solved all the problems the big
bang theory come across all at once, except for the singularity
problem, which is principally impossible to overcome under the
framework of classically general relativity. There seems no avoiding
the singularity problem, and only the quantum cosmology theory
declares to possess the power to solve the diagnose.\\
\subsection{“First three Minutes”in Fine Cosmology}
The time scale is generated together with the Creation of Big Bang.
When time reaches $10^{-43}$ second, the scale grows to
$10^{-33}cm$, the four finds of elementary forces are unified as a
grand force. Only energy lives, without any matter, and the world is
dominated by quantum uncertainty principle. This era is called the
Plank era or the Grand Unified era\cite{standard model}.\\

when it goes to $10^{-36}s$, temperature drops to $10^{29}K$, and
certain kinds of particles are produced. Although some particles
annihilate with their anti-particle companies, the creating rate is
faster than the annihilate rate. Temperature slowing down,
gravitation and the nuclear strong force decouple from the grand
unified force one after one.\\

Then since  $10^{-35}\sim 10^{-33}s$ ,it comes the inflation era.
The spatial scale doubles every $10^{-34}$ second. In a period of
$10^{-32}s$, the scale expanded by $10^{100}$ times, ending with a
spatial scale of $3\times 10^{-25}cm$ and temperature of $3\times
10^{28}K$. By the end of inflation, the electroweak force
decouples into electromagnetic force and nuclear weak force.\\

When it comes to $10^{-11}$th second, the corresponding temperature
is $10^{15}K$. The highest energy  produced on the large hadron
colliders is able to simulate this situation. And the validity of
the numerical calculating and experimental simulating stands testing
and win a wide agreement.\\

When it comes to $10^{-6}$ second, the temperature decreases to
$10^{12}K$, and "quark-gluon soup" is generated. Quarks and gluons
are in free state originally. With the temperature continuously
doing down, a gluon will bound three quarks together to form a
nucleon (proton or neutron). \\

When it comes to 1st second, temperature drops to 10 billion Kelvin.
The contains of the cosmos are photons, quarks, electrons, neutrinos
and other particles, and of course, proton and neutrons. Temperature
decreasing, neutrons begin to decay into steady proton, ending with
a proton-neutron number ratio $7:1$.\\

With time slipping down, the creation rate drops, and annihilation
rate of particle-antiparticle pairs exceed that of creation. Large
amounts of electron-positrons annihilate to produce more photons.
Eventually the annihilation era finished, the bulk being the
remaining quarks and electrons (leptons). The cosmos evolves into a
new creating era, and above all more protons and neutrons are
created.\\

At about 100 seconds after the big bang, it gets 1 billion Kelvin.
Protons and neutrons can never escape from the constraints of
nuclear strong force and forms $H^2_1$. Then $H^2_1$ turns to
${He}^3_2$, and ${He}^3_2$ to ${He}^4_2$, with large
quantities of neutrinos generated meanwhile.\\

Three minutes after the big bang, the light nuclear synthesis
process ends due to low pressure. Now about $25\%$ of protons and
neutrons forms Helium, with a few $H^2_1$, $Li_3$, $Be_4$, and the
remaining neutrons decay into proton. This results in a universe of
$75\%$Hydrogen and $25\%$ Helium.\\

The standard model of particle physics also believes that, in the
extremely early era, some other particles are produced, such as
x-boson, Higgs boson, Weakly Interacting heavy particle, cosmic
string and monomagneton .But none of these are explored. At three
minutes after the big bang, the universe is filled with electron,
nude hydrogen, nude helium and dark matter, as well as high energy
$\gamma$. Because photons, which is the medium particle of
electromagnetic force, interact with free electrons and positrons
via collision, absorption, emitting frequently, the lifetime and
free-diatance of a photon is quite short, and information couldnot
travel to the distance. So, the whole cosmos is in dark plasma
state.\\

0.3 million years after the big bang, it has reached a temperature
as low as 4000K. Electrons are bounded in atoms, so photons could
travel to distant places to compare notes. The cosmos is apparent
now, and it expands steadily and silently, and enriching clustering
structures are formed. The temperature of background radiation goes
on falling ,to the present-day value of 2.7K.
\subsection{Dark Matter}
Just before the discovery of dark energy, the so-called \emph{dark
matter} is already a remarkable challenge to man's brains. Although
Zwicky had suggested the existence of dark matter, it is not until
the late 1970 did the idea is highly treasured. Now, it is commonly
accepted the existence of dark matter. Lots of experiments lead to
this result, the classical representative of which comes from the
measured rotating speed curve of spinning galaxies. With the hard
efforts of high energy physists, conclusions are drawn that the dark
matter, which has hardly any interactions with photons, couldnot be
baryonic. Hence it is now generally called cold dark matter
(hereafter CDM).\\

There are several candidates of elementary particles for CDM. The
most hopeful one is the weakly interacting massive particles (WIMP),
whose mass is probably far beyond a proton. Among several WIMPs, a
particular kind is the neutralino, with a typical mass of 102GeV, or
$10^{-22}$ gram. Other candidates include the axion with a typical
mass of $10^{-6}$eV or $10^{-39}$ gram, fuzzy CDM, neutrinos, etc\cite{dark-matter}.\\

Nevertheless, CDM might also be ancient black wholes that form
before nuclearsynthesis, rather than elementary particles. And the
lower mass limit of the blackholes is Hawking evaporating limit,
namely $10^{15}$ gram.

\section{General Relativity $\&$ Cosmic Dynamics}

\subsection{Gravitational Field Equation $\&$ FRW metric}
The research into cosmic behaviors on large scales is based on
Einstein's gravitational field equation\cite{GR theory1} \cite{GR
theory2}:
\begin{equation}\label{eq:Einstein}
G_{\mu\nu}=R_{\mu\nu}-\frac{1}{2}g_{\mu\nu}R=8\pi GT_{\mu\nu}
\end{equation}
where natural units are employed to set $c=h=1$,and $G_{\mu\nu}$ is
Einstein tesor, $R_{\mu\nu}$ being
Ricci tensor,$R$ being Ricci scalar,$G$ being Newton's gravitational constant.\\

Take another consideration, as for the four-dimensional spacetime
that obeys the Cosmological Principle and is therefore homogeneous
and isotropic, the metric is Friedmann-Robertson-Walker(FRW) metric:
\begin{equation}\label{eq:FRW}
ds^2=dt^2-a^2(t)\left(\frac{dr^2}{1-kr^2}+r^2d\theta^2+r^2\sin^2\theta
d\phi^2\right)
\end{equation}
where $a(t)$ is the scale factor whose evolution is dependent on
cosmic time $t$; $r,\theta,\phi$ is the set of comoving coordinates
fixed on the medium, the medium expanding with the universe while
the comoving coordinates remaining static. $k$ refers to the
curvature scalar. When $k>0$, the spatial topology is limited curved
and namely closed universe; when $k=0$, the spatial topology is
infinitely flat and namely flat universe; when $k<0$, the spatial
topology is infinitely curved and namely open universe. Moreover, if
the unit for $r$ is properly selected, the values of $k$ can be
reduced to be $k=+1,0,-1$ correspondingly. Sometimes, FRW metric
(\ref{eq:FRW}) is transformed to take the formalism below for
computing convenience:
\begin{equation}
ds^2=dt^2-a^2(t)\left[d\chi^2+f^2_k(\chi)(d\theta^2+\sin^2\theta
d\phi^2) \right]
\end{equation}
where
\begin{equation}
f_k(\chi)=\begin{cases} \sin\chi,\qquad&\mbox{$k=+1$}\\
\chi,\qquad&\mbox{k=0}\\
\sinh\chi,\qquad&\mbox{k=\quad -1}
\end{cases}
\end{equation}
\subsection{Perfect Fluid Scenario $\&$ Conservation Laws}
Now let's combine the cosmological principle and the energy-momentum
tensor which reflects the general states of matters, and get:
\begin{equation}\label{eq:ideal liquid}
T_{\mu\nu}=(\rho+p)U_\mu U_\nu+pg_{\mu\nu}
\end{equation}
where the energy density $\rho$ and pressure $p$ are dependent on
$t$ but independent of $r,\theta,\phi$, and $U_\mu$ is the
four-dimensional velocity vector:
\begin{equation}
U^t=1
\end{equation}
\begin{equation}
 U^i=0
\end{equation}
Nevertheless, the energy-momentum tensor for 4D perfect fluid also
takes the form of Eq.(\ref{eq:ideal liquid}), where $\rho$ and $p$
are measured in an inertial reference comoving with the fluid. Hence
perfect fluid scenarios are generally employed as dynamical
mechanism in cosmology.\\

The spatial media(taken as perfect fluid) are pretty homogeneous and
isotropic, which determines the diagonal matrix of its
energy-momentum tensor and the identification of the spatial
elements.
\begin{equation}
T^\mu_\nu=diag(-\rho(t),p(t),p(t),p(t)),
\end{equation}
The total energy-momenata ought to be conserving, hence there being
the covariance equation:
\begin{equation}\label{eq:ener-momen-tensor-conserve}
T^{\mu\nu}_{;\nu}=0
\end{equation}
as to $\nu=r,\theta,\phi$ the above equation holds obviously; as for
$\nu=0$,one gets:
\begin{equation}
d[a^3(\rho+p)]=a^3dp
\end{equation}
and translate it into the formalism of the first law of
thermodynamics:
\begin{equation}\label{eq:thermal1}
d(\rho a^3)=-pd(a^3)
\end{equation}
which indicates,for a comoving volume element, the multiplication of
its pressure and volume (i.e. volume work) $pd(a^3)$ is the opposite
value of the energy change of the mass element $d(\rho a^3)$.
\subsection{Friedmann Equation}
The expanding rate of the universe is reflected via Hubble parameter
$H$, defined as:
\begin{equation}
H\equiv\frac{\dot{a}}{a}
\end{equation}
Hubble parameter evolves according to observations, and is
dynamically treated in theory too. The so-called Hubble Constant
$H_0$ refers to the presentday value of Hubble parameter in
particular.\\

Under FRW meric,the non-null component for the Ricci tensor of
gravitational field equation (\ref{eq:Einstein}) is:
\begin{equation}
R_{00}=-3\frac{\ddot{a}}{a}
\end{equation}
\begin{equation}
R_{ij}=-\left(\frac{\ddot{a}}{a}+2\frac{\dot{a}^2}{a^2}+2\frac{k}{a^2}
\right)g_{ij}
\end{equation}
Ricci scalar being:
\begin{equation}
R=-6\left(\frac{\ddot{a}}{a}+\frac{\dot{a}^2}{a^2}+\frac{k}{a^2}
\right)
\end{equation}
the $0-0$(time) component of the gravitation equation being
\begin{equation}\label{eq:Friedmann}
\frac{\dot{a}^2}{a^2}+\frac{k}{a^2}=\frac{8\pi G\rho}{3}
\end{equation}
This is the very Friedmann equation that represents the dynamical
evolution of cosmic scalar factor, which indicates that that the
expansion process depends on both energy term and geometrical
curvature term. The $i-i$ (spatial) component of gravitational
equation (\ref{eq:Einstein}) is:
\begin{equation}\label{eq:Friedmann2}
2\frac{\ddot{a}}{a}+\frac{\dot{a}^2}{a^2}+\frac{k}{a^2}=-8\pi Gp
\end{equation}
Meantime it is also induced to get Raychaudhuri equation which shows
the evolution of Hubble parameter:
\begin{equation}
\dot{H}=-4\pi G(p+\rho)+\frac{k}{a^2}
\end{equation}
Eq(\ref{eq:thermal1}), Eq(\ref{eq:Friedmann}),
Eq(\ref{eq:Friedmann2}) is connected via Bianchi identity, hence
only two ones are independent. From
(\ref{eq:ener-momen-tensor-conserve}), (\ref{eq:Friedmann}),
(\ref{eq:Friedmann2}) one could induce the energy-momenta conserving
equation for cosmological perfect fluid:
\begin{equation}\label{eq:ener-monen-conserve}
\dot{\rho}+3H(\rho+p)=0
\end{equation}
Unite (\ref{eq:Friedmann})(\ref{eq:Friedmann2})to deduct
$\frac{k}{a^2}$ and one would get a straight overlook on the cosmic
expansion:
\begin{equation}\label{eq:accelerate}
\frac{\ddot{a}}{a}=-\frac{4\pi G}{3}(\rho+3p)
\end{equation}
Conclusion is directly drawn that $\rho+3p<0$ holds for accelerating expansion.\\

Via Hubble parameter, Fredmann equation could be rewritten as:
\begin{equation}
\frac{k}{H^2a^2}=\frac{\rho}{\frac{3H^2}{8\pi G}}-1
\end{equation}
and therefore the cosmic curvature completely depends on the total
energy density, which is in casuality agreement with general
relativity. If the density is above $\frac{3H^2}{8\pi G}$,$k$ will
be positive, and results in a limited and closed cosmos;  If the
density is identical to $\frac{3H^2}{8\pi G}$,$k$ will be null, and
results in a infinite and flat cosmos;  If the density is below
$\frac{3H^2}{8\pi G}$,$k$ will be negative, and results in a
infinite and open cosmos. So in further discussion it is proper to
define the unitary quantity $\frac{3H^2}{8\pi G}$, which share the
dimension of energy density, to be the critical energy density of
the universe:
\begin{equation}
\rho_{crit}\equiv \frac{3H^2}{8\pi G}
\end{equation}
together with a non-dimensional quantity of relative energy density:
\begin{equation}
\Omega\equiv\frac{\rho}{\rho_{crit}}
\end{equation}
According to the current theories, it is $\rho_{crit}$ and $\Omega$
that determine the ultimate fate of the universe. Hence Friedmann
equation take the further form:
\begin{equation}
\frac{k}{H^2a^2}=\Omega-1
\end{equation}
thus a clear graph is finished to reflect how the evolution of the
universe depends on its property parameters, which prepares a firm
foundation for cosmic observations and cosmological dynamics.
\subsection{EOS in Perfect Fluid Scenario}
Actually to resolve cosmological problems under perfect fluid
scenario, we still need the equation of state (EOS):
\begin{equation}\label{eq:state}
p=p(\rho)
\end{equation}
Thus theoretically, the combination of Friedmann Equation
(\ref{eq:Friedmann}), energy-momenta conservation equation
(\ref{eq:ener-monen-conserve}) and EOS (\ref{eq:state}) is equal to
solve the dynamics of cosmos.\\

Generally, a simplest EOS is employed for cosmological perfect
fluid:
\begin{equation}
p=w \rho
\end{equation}
where $w\in\mathbb{R}$. For ordinary matter, $w=0$. And the
accelerating expansion conditions $\rho+3p<0$ can be re-expressed
via EOS parameter:
\begin{equation}
w <-1/3
\end{equation}\\
Integration by (\ref{eq:thermal1}) give rise to the relation of
energy density and cosmic scalar factor:
\begin{equation}
\rho\varpropto a^{-3(1+\omega)}
\end{equation}
When the universe is dominated by relativistic matter and radiation
(i.e. the baby era), $w=1/3$, hence
\begin{equation}
a(t)\propto (t-t_0)^{1/2},\qquad \rho\propto a^{-4}
\end{equation}
When the universe is dominated by non-relativistic matter, $w=0$:
\begin{equation}
a(t)\propto (t-t_0)^{2/3},\qquad \rho\propto a^{-3}
\end{equation}
All in all, different eras correspond to different domination and
evolution.

\chapter{Accelerating Expansion of the Universe}
\section{Accelerating Expansion and Dark Energy}
The inflation theory, which is a great success as supplement to Big
Bang cosmology, declares that the cosmos is flat in early 1980s.
Theorists' insistence lead to annoying disagreement. However, cosmic
observations found that there is far less matter than needed for
flat topology. Shortly afterwards in early 1990s, COBE is sent by
NASA to explore the cosmic microwave background. To theorists'
delight, the location of the first peak in angular power spectrum
confirms the flatness of the cosmos. This strongly require a new
density component to fit the remarkable gap between the matter
density and the critical density. Before physists could take a calm
breath, the redshift measurement of \textrm{1}a supernova declared
the accelerating expansion of the universe! This milestone is among
the greatest ever discoveries, and bring the discoverers the Shaw
Awards
for Astronomy.\\

So, CMB exploring proves the flatness, homogeneity and isotropy of
the universe, and SN$1a$ confirms the accelerating expansion. These
results lead to nondebatable conclusions that, if general relativity
and the standard model of particle physics are correct, then\\

(1) The density of cosmos is $\Omega=1$. Since baryons, non-baryonic
CDMs and radiation add up to only around
$\Omega_m+\Omega_\gamma=0.3$, the major energy (the gap) comprise of
an
unknown component, dubbed Dark Energy (DE).\\

(2) Only limited dynamical characters are known. Dark energy has
positive energy, negative pressure, and anti-gravitation effects
that drive the accelerating expansion. DE doesnot condensate at
scales smaller than 100Mpc; DE density is almost constant; DE is
only dominant recently ($z\in (0\sim 1)$ and $z_0=0$ today), so it
doesn't bother the nuclear-synthesis process at early era.\\

And even up till now, we are still quite unfamiliar with DE. DE may
be scalar fields, vector fields, tensor fields, or even cosmic
strings, etc\cite{DavSper}. And a set of typical data is that the
critical density is made up of $72\%$ DE, $24\%$ CDM, $4\%$ ordinary
matter.
\section{Characteristic Parameters of Cosmos}
The properties of the universe need to be parameterized for
quantitative study (this is a common method to handle huge and
complicated systems).\\

Presently, there are 12 parameters that are mostly
popular.\cite{quintom3}:
$$P\equiv (\omega_b,\quad \omega_c,\quad \mathcal{K},\quad H,\tau, \quad
w_0,\quad w_1, \sum m_\nu,\quad n_s,\quad A_s,\quad \alpha_s,\quad
r)$$ where $\omega_b\equiv\Omega_bh^2$ and
$\omega_c\equiv\Omega_ch^2$ represent the relative energy density of
baryonic matter and cold dark matter. $\mathcal{K}$ represents the
presentday scalar curvature ,and cosmic observations support the
result of $\mathcal{K}=0$ ,a flat cosmos；$H$ represents Hubble
parameter;；$\tau$ represents luminous depth；$\sum m_\nu$
represents the total mass of the three generation of neutrinos. . In
general computing process, the EOS parameter is often rewritten as
$w=w_0+w_1(1-a)$,$w_0$ and $w_1$ represent the new EOSs
respectively. The last foue parameters are related to inflation
theory.$A_s$ represents the amplitudes of original power；$n_s$
represents the power index for scalar spectra；$\alpha_s$
represents shift of spectra index $n_s$ ；$r$ is the tensor power parameter.\\

It is of great importance to investigate these patameters , which
help constraint the properties of dark energy. Here below we list
the data from Willkinson Microwave Anisotropy Probe, which can de
downloaded from the WMAP homepage of NASA.
\section{Mechanism for Accelerating Expansion}
Ever since the proof that DE is dominant, its components, origin,
dynamics are waiting to be determined. With the parameter
constraining via cosmic observations, various models have been put
forward for a cosmos with accelerating expansion. I would divide
these models into three categories for further discussion.

\newpage
Table: Data from WMAP.
\begin{center}
\renewcommand\arraystretch{1.3}
\begin{tabular}{|c||cccc|}
 \hline
Cosmic Parameters     &  Symbol  &  Value  & + uncertainty    & -uncertainty\\
\hline \hline
total density  &$\Omega_{tot}$ &   1.02   &    0.02       &   0.02\\

Quintessence EOS   & $\omega$  & $<-0.78$  & $95\% CL$  & --  \\

DE density   &$\Omega_\Lambda$  & 0.73  &0.04   & 0.04  \\

Baryon Energy Density  & $\Omega_b$  & 0.044  & 0.004  & 0.004\\

Baryon Number Density($cm^{-3}$)  & $n_b$ & $2.5\times 10^{-7}$ & $0.1\times 10^{-7}$  & $0.1\times 10^{-7}$ \\

Matter Energy Density & $\Omega_m$  & 0.27  & 0.04 & 0.04 \\

Neutrino Density & $\Omega_\nu h^2$  & $<0.0076$ & $95\% CL$  & -- \\

CMB Temperature(K)  & $T_{cmb}$ & 0.275  & 0.002 & 0.002 \\

Baryon-photon Number  & $\eta$ & $6.1\times 10^{-10}$ & $0.3\times 10^{-10}$& $0.2\times 10^{-10}$ \\

$8h^{-1}Mpc$Spherical Fluctuation.  & $\delta_8$ & 0.84 & 0.04 & 0.04\\

Scalar Spectra Index$k_0=0.05Mpc^{-1}$) & $n_s$ & 0.93 & 0.03 & 0.03 \\

Index shift($k_0=0.05Mpc^{-1}$) & $dn_s/dlnk$ & -0.031 & 0.016 & 0.018\\

Scalar-Tensor Ratio($k_0=0.002Mpc^{-1}$) & r & $<90\%$  & $95\% CL$ & -- \\

decoupling redshift & $z_{dec}$ & 1089 & 1 & 1 \\

decoupling width(FWHM) & $\Delta z_{dec}$ & 195 & 2 & 2 \\

Hubble parameter(reduced) & h & 0.71 & 0.04 & 0.03 \\

Cosmos Age(Gyr) & $t_0$  & 13.7 &0.02 & 0.02\\

Decoupling time(Kyr) & $t_{dec}$ & 379 & 8 & 7\\

Heavy Ionic time(Myr,$95\% CL$) & $t_r$ & 180 &220 &2580\\

decoupling interval(Kyr) & $\Delta t_{dec}$ & 118 &3 &2 \\

equality redshift & $z_{eq}$ & 3233 & 194 & 210\\

Baryonic luminous depth & $\tau$ & 0.17 &0.04 &0.04 \\
\hline

\end{tabular}
\end{center}

\subsection{Scalar Fields}
Bulk of these proposals are some kinds of scalar fields with shallow
potentials and thus tiny mass. The field rolls slowly, up
(canonical) or down (noncanonical), in its own potential. This
process takes place only considerable on Hubble time scale. EOS in
scalar field models takes the traditionally most simple form
$p=w\rho$, and EOS parameter $w$ turns the identity of different
fields. $w>-1$ or $-1<w<-1/3$ means Quintessence field; $w=-1$ is
Einstein's cosmological constant $\Lambda$; $w<-1$ reflects Phantom
field. These cases are all real fields; and those with $w$ crossing
$-1$ can be real (Quintom) or complex (Hessence).\\

And the key point is that a scalar field is itself meaningful at
all, if not treated together with the accelerating expansion.
Firstly, it's generated or associates with many attractive theories,
such as superstring quantum field, pseudo-Nambu-Goldstone Bosonic
Formulation, Brans-Dicke theory. So it's a natural trial to test the
cosmological consequence of a scalar field, typical work like Ratra
and Peebles in late 1980s. Secondly, a scalar field could help make
clear the CDM spectra. Last but not least, even a light scalar field
can generate testable consequences in the standard CDM scenario. For
those reasons, scalar fields such as Quintessence has been discussed
pioneeringly, after the inflation theory, before the accelerating
expansion observations. More discussions are available in this thesis afterwards. \\

\subsection{Modified Gravitational Field Equation/Friedmann Equation}
A most remarkable candidate is Cardassian models, which add a new
energy term to the right side of Friedmann equation, or to the
energy-momenta tensor of Gravitational Field Equation. This
Cardassian term depends on the density of matter and radiation only,
and the modified Friedmann equation could produce a flat, matter
dominated universe with accelerating expansion at late time, while
the early evolution of baby cosmos is not perturbed. The creative
work of this thesis is exactly connected with this models. \\

Modifications to gravitation equation includes operations on the
geometric term of Ricci tensor, or energy-momenta tensor. Besides
Cardassian, other examples include Randall-Sundrum model, etc.
\subsection{Modern QFT approaches}
Such models include k-essence, tachyon and Chaplygin gas. K-essence
focuses on modifying the kinetic energy term of Lagrangian, who acts
as the first principle. Tachyon plays an important role in standard
model of particle physics and superstring guage theory, and its
connections with accelerating expansion is established after 1998.
Both k-essence and tachyon modeling are deeply rooted in quantum
field theory.\\

Chaplygin gas seems less mathematically complicated than k-essence
and tachyon formulation. It is a kind of non-relativistic perfect
fluid in superstring theory, being the only liquid allowing
supersymmetric generalization. Its EOS is quite different from that
of scalar fields, and can be reached via several approaches, such as
non-relativistic approximation of Born-Infield theory. In
(generalized) Chaplygin gas cosmology, the ultimate fate of
accelerating expansion is de Sitter Universe. A most interesting
property is that (generalized) Chaplygin gas indicates a unified
description of dark energy with dark matter.\\

In the following test, I firstly review some typical existing work,
including four scalar fields (Quintessence, Phantom, Quintom,
Hessence), k-essence and Chaplygin gas. I put my work in the last
two chapters, where I come up with a new Cardassian model and
introduce virial collapse to Cardassian cosmology.

\chapter{Quintessence \&  Phantom   }

\section{Quintessence Scalar Field}
Quintessence scalar field is the first introduced dark energy model,
and has received most attention. At early 1980s, a scalar field is
employed in the inflation theory, which encourages people to analyze
the consequence of a scalar field rolling in its own potential at
the whole Hubble time scales, rather than the tiny time interval at
the extremely early universe. This work is done in late 1980s, and
is re-introduced as the first non-$\Lambda$ mechanism for dark
energy after 1998.
\subsection{Basic Structure of Quintessence}

Generally Quintessence is treated as a classical scalar field, with
smallest coupling with gravitation field, and its action is:
\begin{equation}\label{eq:quint-action}
S=\int \mathcal {L}_\phi \sqrt{-g}d^4x
\end{equation}
where $g$ is the determinant of the matrix $g_{\mu\nu}$. With metric
$(+ - - -)$,phenomenently,the Lagrangian density with positive
energy and positive pressure $\mathcal{L}_\phi$ is:
\begin{equation}
\mathcal{L}_\phi=\frac{1}{2}g^{\mu\nu}\partial_\mu\phi\partial_\nu\phi-V(\phi)
\end{equation}
Take variance operation on (\ref{eq:quint-action}), and one gets the
energy-momenta tensor for a scalar field:
\begin{equation}
T^\phi_{\mu\nu}=\partial_\mu\phi\partial_\nu\phi-g_{\mu\nu}\left[\frac{1}{2}\partial_\mu\phi\partial_\nu\phi-V(\phi)\right]
\end{equation}
Take FRW metric(\ref{eq:FRW}) in a flat space into account, via the
conservation equation $T^{\mu\nu}_{\phi\quad;\nu}=0$, one can get
the dynamical equation for the scalar field $\phi$:
\begin{equation}
\ddot{\phi}+3H\dot{\phi}+\frac{dV}{d\phi}=0
\end{equation}
Assuming that the scalar field is homogeneous, whose derivative to
spatial coordinates is zero and dependent on cosmological time only.
Via FRW line elements in such a field, one can get the energy
density  and pressure of Quintessence field:
\begin{equation}
\rho_\phi=\frac{1}{2}\dot{\phi}^2+V(\phi)
\end{equation}
and
\begin{equation}
p_\phi=\frac{1}{2}\dot{\phi}^2-V(\phi)
\end{equation}
Hence EOS parameter of Quintessence field is:
\begin{equation}\label{eq:quint-state}
w_\phi=\frac{p_\phi}{\rho_\phi}=\frac{\frac{1}{2}\dot{\phi}^2-V(\phi)}{\frac{1}{2}\dot{\phi}^2+V(\phi)}
\end{equation}
The dynamical equations for the evolutionary cosmos is
\begin{equation}
H^2=\frac{8\pi G}{3}\left[\frac{1}{2}\dot{\phi}^2+V(\phi)  \right]
\end{equation}
\begin{equation}\label{eq:quint-accelerate}
\frac{\ddot{a}}{a}=-\frac{8\pi G}{3}\left[\dot{\phi}^2-V(\phi)
\right]
\end{equation}
From Eq.(\ref{eq:quint-accelerate})indicates that accelerating
expansion is possible if $\dot{\phi}^2<V(\phi)$ holds. Insert
(\ref{eq:quint-state}) and (\ref{eq:ener-monen-conserve}) into
Quintessence field, then:
\begin{equation}
\rho_\phi=\rho_0exp\left[-\int 3(1+w_\phi)\frac{da}{a} \right]
\end{equation}
where $\rho_0$ is an integral constant. According to
(\ref{eq:quint-accelerate}), the value of $w_\phi$ must be mamong -1
to 1. If the rolling in the potential proceeds very slowly, i.e.
$\dot{\phi}^2\ll V(\phi)$,then $\omega_\phi=-1$ and $\rho_\phi$
being a constant; If the rolling in the potential proceeds very
sharply, i.e. $\dot{\phi}^2\gg V(\phi)$,then $\omega_\phi=1$ and
$\rho\propto a^{-6}$；As to the media cases, Quintessence energy
density will be
\begin{equation}
\rho_\phi\propto a^{-m},\qquad 0<m<6
\end{equation}
Since $\omega=-1/3$ is the critical condition for accelerating
expansion and decelerating expansion, constraints can be
strengthened for $0<m<2$ for accelerating expansion.
\subsection{Quintessence Potential}
In discussions with Quintessence field, three potentials are most
popular\\:

(1)Inverse power style potential by supersymmetry
models\cite{quint-string1}:
\begin{equation}
V(\phi)=V_0\left(\frac{\phi_0}{\phi}  \right)^{a_0}\qquad a_0>1
\end{equation}
(2)Exponential potential by Kaluza-Klein theory\cite{quint-string2}:
\begin{equation}
V(\phi)=V_0e^{\phi-\phi_0}
\end{equation}
(3)Cosine potential by Pseudo-Nambu-Goldstone Bosonic
Mechanism\cite{quint-string3}:
 \begin{equation}
V(\phi)=\frac{1}{2}V_0\left[\cos(\frac{\phi-\phi_0}{f} )+1 \right]
\end{equation}\\
Let's take a closer look through a concrete sample: For power style
expansion, the driven potential is exponential. The scalar factor
expands in power style, $a(t)\propto t^p$, where $p=1$ is the
critical condition to separate accelerating and decelerating
expansion. According to (\ref{eq:Friedmann2}), one has
\begin{equation}
\dot{H}=-4\pi G\dot{\phi}^2
\end{equation}
So one could employ the observables $H$ and $\dot{H}$ to represent
the potential $V(\phi)$ and $\dot{\phi}$:
\begin{equation}
V=\frac{3H^2}{8\pi G}\left(1+\frac{\dot{H}}{3H^2} \right)
\end{equation}
\begin{equation}
\phi=\int dt\left(-\frac{\dot{H}}{4\pi G} \right)^{1/2}
\end{equation}
This operation can be treated as the reconstruction of the model.
Hence the driven potential for a power style expanding potential is:
\begin{equation}
V(\phi)=V_0exp\left(-\sqrt{\frac{16\pi}{p}}\frac{\phi}{m_{pl}}
\right)
\end{equation}
where $V_{0}$ is a constant, unitary parameter $m_{pl}$ being Plank
mass. In a word, an exponential Quintessence potential will drive
the universe expanding in power style. Moreover, exponential
Quintessence could lead to scaling solutions of dynamical phase
system, where the ratio of Quintessence energy density $\rho_\phi$
to that of background matter $\rho_m$ is constant.
\subsection{Scalar Fields' problems and Tracker Field}
If one employs a slowly rolling and light scalar field to drive the
accelerating expansion, two serious problem will arise. This is the
difficulty that all scalar field come across.\\

(1)Fine-tuning Problem\\
The kernel of this problem lies that, why the missing energy density
differs so much in comparison with the typical energy scale in
particle physics ? According to thee current data, $\Omega_m=0.3$,
then dark energy density is in the magnitude of $10^{-47}GeV^4$,
which needs an extra 14 magnitude beyond electromagnetic
interactions. \\

(2)Coincidence Problem\\
The kernel of this problem lies at the initial conditions of the
cosmos. The evolutionary rate of Quintessce field differs very much
with that of energy density, and the process is quite complicated.
So, to evolve into the state where the energy density of
Quintessence field and matter are of the same magnitude after
137Gyr, the initial conditions must be exactly set.\\

All scalar fields cosmological models are upset with the two
problems, but, the discovery of the so-called Tracker Field, which
originate from Quintessence and then spread to all other scalar
fields, throws light on beating these
problems\cite{quint-tractor-field}. It's found that, as for some
special potential, such as $V\sim V_0\phi^{-n}$, $V\sim
V_0\exp(M/\phi-1)$, $V\sim
V_0[sinh(\alpha\sqrt{k_0}\Delta\Phi)]^\beta$ \cite{quint-tracker-2},
there are tracker solution for the Quintessence scalar field.
Tracker solutions have quite unrestricted initial conditions, and
the energy density ratio of Quintessence to matter is permitted to
fluctuate for 100 magnitudes to have the same evolutionary results.
This means it's almost independent of initial conditions, or
perfectly Markovian. This solves the two problems referred to above
in some sense. And from then on, the existence of tracker solution,
in the differential autonomous dynamical system  turn an important
judgement for the quality of a scalar field for dark energy, and for
the quality of a potential for certain scalar field.
\subsection{Typical discussions for Quintessence}
(1)In \cite{quint-constraint potential} it's shown that the
gravitational field equation and conservation laws put strict
constraints to Quintessence with $w=const.$, For constant EOS
parameter $w$ and $w\neq-1$ (simply the case of cosmological
constant $\Lambda$):
\begin{equation}
\frac{V'}{V'_0}=\sqrt{\Omega_\phi\left(\frac{V}{V_0}
\right)^2+\Omega_M\left(\frac{V}{V_0}
\right)^{\frac{w+2}{w+1}}+\Omega_k\left(\frac{V}{V_0}
\right)^{\frac{3w+5}{3w+3}}}
\end{equation}
where V is the scalar potential for Quintessence field,
 $V'\equiv\pm3H_0\sqrt{(1-w^2)V_0/2}$,$V_0$
being the presentday value.\\

(2)Carroll made typical evaluation to the strength of Quintessence
field\cite{quint-typical-2}. Cosmological observations restrict that
the scalar potential is quite shallow, which means the tiny mass in
excited states, $m_\phi\equiv\sqrt{V''(\phi)/2}\leq H_0\sim
10^{-33}eV$. In order to support the observed energy density, the
current value of the potential must be approximately the closure
density, $V(\phi_0)\sim(10^{-3}eV)^4$, and the field be
$\phi_0\sim10^{18}GeV\sim M_{pl}=(8\pi G)^{-1/2}$, where $M_{pl}$ is
the reduced Plank mass. A light scalar field will generate testable
long-range force, and if Quintessence could couple with matter, it
would lead to the evolution of physical constants at Hubble time
scale.\\

(3)\cite{quint-typical-3} investigates the origin of Quintessence
from breakdown of supersymmetry. \cite{quint-typical-3-2}
investigated the possibility that axions and QCD massless quark
comprise the Quintessence. As a matter of fact, in the following
years since the discovery of dark energy, attentions are focused on
the theoretical formulation, and practical constrains for
Quintessence takes fall behind obviously. Various connections with
Quitessence from other branches have been established, and large
amounts of theses are available, but this thesis wouldnot go further
and wider review for Quintessence.
\section{Phantom Scalar Field}
According to Riess et al, if the matter density is treated as
$\Omega^0_m=0.27\pm 0.04$, one would see $w=-1.02^{+0.13}_{-0.19}$,
and $-1.46<w<-0.78$ for $95\%$ confidence level. If one goes further
to take the results of CMB and 2dFGRs into consideration, it's
$w=-1.08^{+2.0}{-0.18}$. So, $w<-1$ is also possible. Caldwell for
the first time discussed the cosmological consequence for $w<-1$,
and introduced the second scalar field, Phantom\cite{phantom-birth},
as opposed to Quintessence. Phantom has many amazing properties.
Phantom violates the weak energy conditions of gravitation; $w<-1$
leads to the effective sound speed $v=\sqrt{|dp/d\rho|}$ beyond
light velocity. The most debating result is the so-called Big Rip.
The dark energy density is dynamical and increase along with the
accelerating expansion; thus the increasing dark energy and the
accelerating expansion unite to form a positive feedback pair, and
one day in the future, the huge negative pressure will tear out the
spacetime and all matter. For a review, see \cite{big rip-rev} and
\cite{big rip-conditions}.
\subsection{basic Structures of Phantom}
Canonical scalar fields couldnot realize the process for
$\omega<-1$, so noncanonical ones are employed for Phantom. A
noncanonical Lagrangian for free Phantom field, which also has the
smallest couplings with gravitation, can be written
as\cite{phantom-birth}\cite{phantom-model}:
\begin{equation}
\mathcal{L}_\phi=-\frac{1}{2}g^{\mu\nu}\partial_\mu\phi\partial_\nu\phi-V(\phi)
\end{equation}
Hence the dynamical equation for Phantom field is:
\begin{equation}
\ddot{\phi}+3H\dot{\phi}-\frac{dV}{d\phi}=0
\end{equation}
the minus sign before thee potential term makes Phantom climbs up
the potential, as opposed to a canonical field's rolling down in its
potential (see Quintessence).\\

Variance of Phantom Lagrangian leads to Phantom energy density and
pressure:
\begin{equation}
\rho_\phi=-\frac{1}{2}\dot{\phi}^2+V(\phi)
\end{equation}
and
\begin{equation}
p_\phi=-\frac{1}{2}\dot{\phi}^2-V(\phi)
\end{equation}
as well as Phantom EOS (To avoid the increase of Tackyon module, $w$
is always assumed to be constant in Phantom):
\begin{equation}
w_\phi=\frac{p_\phi}{\rho_\phi}=\frac{-\frac{1}{2}\dot{\phi}^2-V(\phi)}{-\frac{1}{2}\dot{\phi}^2+V(\phi)}
\end{equation}
Obviously $w_\phi<-1$ when $\frac{1}{2}\dot{\phi}^2<V(\phi)$. And
energy-momenta tensor for Phantom field is:
\begin{equation}
T^\phi_{\mu\nu}=-\partial_\mu\phi\partial_\nu\phi-g_{\mu\nu}\left[\frac{1}{2}\partial_\mu\phi\partial_\nu\phi-V(\phi)\right]
\end{equation}
\subsection{Hubble Parameter \& Phantom Cosmology}
The energy density of matter becomes equal with that of Phantom
field at the cosmological time $t_{eq}$, and the scalar factor
driven by Phantom energy will be:
\begin{equation}
a(t)\simeq a(t_{eq})\left[(1+w)\frac{t}{t_{eq}}-w
\right]^{\frac{2}{3(1+w)}},\quad w<-1
\end{equation}
So the scalar factor will be divergent one day, i.e.$t\rightarrow
t_{BR}=\left(\frac{w}{1+w} \right)t_{eq}$ and
$a(t)\rightarrow\infty$. Insert $w<-1$ into Friedmann Equation:
\begin{equation}
H(z)=H_0\left[\Omega_m(1+z)^3+\Omega_X(1+z)^{3(1+w)} \right]^{1/2}
\end{equation}
where $H_0=H(z=0)$ means the presentday value of Hubble parameter,
and $w=w_{DE}$ refers to EOS parameter of dark energy.
$\Omega_m=\frac{8\pi G\rho_{0m}}{3H^2_0}$, $\Omega_X=\frac{8\pi
G\rho_{DE}}{3H^2_0}$, is relative energy density for matter and dark
energy respectively, and $\Omega_m+\Omega_X=1$. Thus,
\begin{equation}
H(z)\sim t^{-1}_{eq}\left[(1+w)\frac{t}{t_{eq}}-w \right]^{-1}
\end{equation}
When $t\rightarrow t_{BR}=\left(\frac{w}{1+w} \right)t_{eq}$, the
Hubble parameter is also divergent mathematically. This indicates
that one day the expanding velocity will approach the extremely
upper limit. So Phantom dark energy density $\rho(t)\propto
\left[\Omega_m(1+z)^3+\Omega_X(1+z)^{3(1+w)} \right]^{-2}$ is
divergent at Big Rip, too. At the Big Rip singularity, spacetime and
matter are all destroyed.\\

Phantom is also a kind of scalar field under large amounts of
discussions, and its connections with quantum field theory, standard
model of particle physics, superstring theory and quantum cosmology
have all been established \cite{phantom-connections-1}
\cite{phantom-connections-2} \cite{phantom-connections-3}
\cite{phantom-connections-4}, etc. And not all Phantom cosmos of
$w_\phi<-1$ ends up with Big Rip, but extra mechanism must be
introduced,see \cite{big rip-avoid-1} \cite{Odintsov1}
\cite{Odintsov2} \cite{big-rip-avoid-2} \cite{big-rip-avoid-3}
\cite{big-rip-avoid-4} \cite{big-rip-avoid-5} and the pioneering
work of classification of Big Rips from Qhantom and Quintessence
theories\cite{Odintsov3}, as well as alternative attempt to quintom
\cite{Odintsov1}.

\chapter{Quintom \& Hessence  }
\section{Quintom Scalar Field}
\subsection{Scalar Field with $w$ crossing -1}
Riess et al analysed their SN1a Gold data (with dark energy EOS
parameter $w(z)=w_0+w'z$), and found $w=-1.02^{+0.13}_{-0.19}$,
which indicates that the presentday EOS parameter is probably
crossing -1. Shortly, Alam, Sahni et al combined Gold data with CMB
data to examine the behaviors of dark energy, and they drew clear
conclusion that EOS parameter evolves continuously, from $w\simeq 0$
at $z\simeq 1$, to $w \leqslant -1 $ at $z\simeq 0$ today.\\

Hubble parameter and luminal distance is connected via:
\begin{equation}
H(z)=[\frac{d}{dz}(\frac{d_L(z)}{1+z})]^{-1}
\end{equation}
Alam et al selected a formalism in which Hubble parameter is
independent of models:
\begin{equation}
H(x)=H_0[\Omega_{m0}x^3+A_0+A_1x+A_2x^2]^{\frac{1}{2}}
\end{equation}
where $x=1+z$, and $\Omega_{m0}$ being the presentday matter
density, and $A_0$, $A_1$, $A_2$ are constants. Thus EOS parameter
is reconstructed tobe:
\begin{equation}
\omega(x)=\frac{(2x/3)H'/H-1}{1-(H_0/H)^2\Omega_{m0}x^3}
\end{equation}
where prime $'$ refers to derivative with $x$. Together with modified $\chi^2$ test, the above
conclusions are drawn.\\

Hunterer confirmed the result of crossing -1, via data fitting of
piecewise function and cosmic obsevations, and to be more exact, the
cosmological redshift $z<0.2$ for $w<-1$, and $z>0.2$ for $z>-1$.
Corasaniti went even further to make combined constraints with CMB,
SN1a, and large scale exploring data, to get the result of $w$
crossing -1.\\

As it is put above, $-1<w<-\frac{1}{3}$ for Quintessence and $w<-1$
for Phantom, and this constraints is principly inviolated, so, both
Quintessence and Phantom are powerless to realize the crossing of
the state $w=-1$. And therefore, the so-called Quintom scalar field
is put forward, which aims at crossing EOS -1.\\

However, to build such a Quintom model, one runs into the
\emph{No-Go} theorem (see A. Vikman\cite{Vikman}): (1)Under the
frame of four dimensional FRW metric and general relativity, (2)if
dark energy is a single perfect fluid, or single scalar field $\phi$
with Lagrangian $\mathcal{L}=\mathcal{L}(\phi,
\partial_\mu\phi\partial^\mu\phi)$, (3)with smallest coupling to
gravitation, then the EOS parameter $w$ couldnot cross -1. This
means, to make $w$ cross -1, at least one of the three conditions
should be violated, such as Quintom with double
field\cite{quintom1}, or Quintom based on Lagrangian with high order
modification term\cite{quintom2}.
\subsection{Basic Structure of Double Field Quintom}
Quintom field includes two real scalar fields, one being canonical
Quintessence style field $\phi$ ,the other being Phantom style field
$\varphi$.\\

Lagrangian deensity of Pantom field is:
\begin{equation}
\mathcal{L}=\frac{1}{2}\partial^\mu\phi_Q\partial_\mu\phi_Q-\frac{1}{2}\partial^\mu\phi_P\partial_\mu\phi_P-V(\phi_Q,\phi_P)
\end{equation}
thus Quitom dark energy density is:
\begin{equation}
\rho_q=\frac{\dot{\phi}^2}{2}-\frac{\dot{\varphi}^2}{2}+V(\phi,\varphi)
\end{equation}
and with pressure of:
\begin{equation}
p_q=\frac{\dot{\phi}^2}{2}-\frac{\dot{\varphi}^2}{2}-V(\phi,\varphi)
\end{equation}
With the dark energy density and pressure, EOS parameter of Phantom
dark energy is:
\begin{equation}\label{eq:quitom-state}
w_q=\frac{p_q}{\rho_q}=\frac{\dot{\phi}^2-\dot{\varphi}^2-2V(\phi,\varphi)}{\dot{\phi}^2-\dot{\varphi}^2+2V(\phi,\varphi)}
\end{equation}
From Eq(\ref{eq:quitom-state}) we can see if
$\dot{\phi}^2>\dot{\varphi}^2$, $\omega
>-1$; if $\dot{\phi}^2<\dot{\varphi}^2$,$\omega
<-1$. EOS of Quintom crosses -1.\\

At first these two componential potential have no interactions with
each other, Quintom potential is the simple addition of those two
potentials:
\begin{equation}
V(\phi,\varphi)=V_1(\phi)+V_2(\varphi)
\end{equation}
take a simple potential $V(\phi_Q,\phi_P)$ for example:
\begin{equation}
V(\phi_Q,\phi_P)=V_0\left[exp(-\frac{\lambda}{M_P}\phi_Q)+exp(-\frac{\lambda}{M_P}\phi_P)
\right]
\end{equation}
For these cases, the steady attractor solution is in Phantom style,
or the ultimate state is Phantom, where one has to face the
nightmare of Big Rip.

\subsection{Generalized Quintom}
There are various methods to make expansion, a simplest and natural
way is to set weights for the addition of Quintessence and Phantom.


\section{Hessence Scalar Field}
Quintessence, Phantom, Quintom are all real, and the earliest
attempt to describe dark energy is Spintessence model. However,
Spintessence falls into deadly trouble when dealing with the
steadiness of Q-balls and is generally abandoned as a dark energy
model(but Spintessence turns out to be a good candidate for dark
matter).
\subsection{Basic Structure of Hessence}
Considering a single complex scalar field with EOS parameter
crossing -1\cite{Hessence1}\cite{Hessence2}:
\begin{equation}
\Phi=\phi_1+i\phi_2
\end{equation}
As for complex field, its canonical Lagrangian is:
\begin{equation}
\mathcal{L}=\frac{1}{2}(\partial_\mu\phi_1)^2+\frac{1}{2}(\partial_\mu\phi_2)^2-V(|\Phi|)
\end{equation}
But, Dark energy based on canonical Lagrangian is not steady
(Spintessence). Thus, the modified noncanonical Lagrangian is
introduced as the first principle for Hessence complex model:
\begin{displaymath}
\mathcal
{L}_{he}=\frac{1}{4}[(\partial_\mu\Phi)^2+(\partial_\mu\Phi^\ast)^2]-U(\Psi^2+\Psi^{\ast
2})
\end{displaymath}
\begin{equation}\label{eq:he-lagrangian}
=\frac{1}{2}[(\partial_\mu\phi)^2-\phi^2(\partial_\mu\theta)^2]-V(\phi)
\end{equation}
the action of Hessence field is:
\begin{equation}
S=\int d^4x\sqrt{-g}\left(-\frac{R}{16\pi
G}+\mathcal{L}_{he}+\mathcal{L}_{m} \right)
\end{equation}
where $(\psi,\theta)$ are two newly defined variables for Hessence:
\begin{equation}
\phi_1=\phi\cosh\theta, \qquad \phi_2=\phi\sinh\theta
\end{equation}
thus
\begin{equation}
\phi^2=\phi_1^2-\phi_2^2, \qquad \coth\theta=\frac{\phi_1}{\phi_2}
\end{equation}
This is why Hessence receives its name, and $H$ here stands for $hyperbolic$.\\

In flat FRW spacetime with scalar factor $a(t)$,  and supposing
$\phi$ and $\theta$ is isotropic, from Hessence Lagrangian
(\ref{eq:he-lagrangian}), one gets Hessence dynamical equation which
depends on $\phi$ and $\theta$:
\begin{equation}
\ddot{\phi}+3H\dot{\phi}+\phi\dot{\theta}^2+\frac{\partial
V}{\partial\phi}=0
\end{equation}
\begin{equation}\label{eq:hessence}
\phi^2\ddot{\theta}+(2\phi\dot{\phi}+3H\phi^2)\dot{\theta}=0
\end{equation}
Thus Hessence energy density and pressure are:
\begin{equation}
\rho_{he}=\frac{1}{2}(\dot{\phi}^2-\phi^2\dot{\theta}^2)+V(\Psi)
\end{equation}
\begin{equation}
p_{he}=\frac{1}{2}(\dot{\phi}^2-\phi^2\dot{\theta}^2)-V(\Psi)
\end{equation}
Based on(\ref{eq:hessence}),there's
\begin{equation}\label{eq:Q}
Q=a^3\Psi^2\dot{\theta}=const.
\end{equation}
where $Q$ is the conservational quantity in certain volume, which is
generated due to inner symmetry of complex scalar fields. In
details, the Hessence Lagrangian (\ref{eq:he-lagrangian}) is
conserving under local gauge transformations, or remain a same
mathematical form after $\phi\rightarrow\phi $ and
$\theta\rightarrow\theta-i\alpha$. According to N$\ddot{o}$ether
theorem, conserving fluid and quantity associates (noted $Q$). On
the other hand, Hessence potential $V(\Phi,\Phi^\ast)$ or
$V(\phi_1,\phi_2)$ should only depend on $\Phi^2+\Phi^{\ast 2}$ or
$\phi_1^2-\phi^2_2$(or noted $V(\phi)$), to keep the symmetry
holding. Now turn attention to the comparison of Hessence Lagrangian
with Quintom Lagrangian, which shows that it is $\theta$ term or $Q$
term that plays the role of Phantom field. When $Q=0$, Hessence
reduces to Quintessence, and there is no more necessity for crossing
$w=-1$. And there's the pity that physical meaning of $Q$ is still
uncertain.\\

It should be stressed that Hessence differs from Quintom and is an
independent model. An important difference is that
$\phi_1^2-\phi_2^2$ is treated as a whole part to be the independent
variable, while the two componential fields are generally
independent to each
other in Phantom.\\

With Eq(\ref{eq:Q}) one can rewrite the dynamical equation, energy
density and pressure of Hessence as:
\begin{equation}\label{eq:he-conserv}
\ddot{\phi}+3H\dot{\phi}+\frac{Q^2}{a^6\phi^3}+\frac{\partial
V}{\partial \phi}=0
\end{equation}
\begin{equation}
\rho_{he}=\frac{1}{2}\dot{\phi}^2-\frac{Q^2}{2a^6\phi^2}+V(\phi)
\end{equation}
\begin{equation}
p_{he}=\frac{1}{2}\dot{\phi}^2-\frac{Q^2}{2a^6\phi^2}-V(\Psi)
\end{equation}
I t is notable that (\ref{eq:he-conserv}) is equal to Hessence
energy conserving equation:
\begin{equation}
\dot{\rho}_{he}+3H(\rho_{he}+p_{he})=0
\end{equation}
while Friedmann and Raychaudhuri equations turn to be
\begin{equation}\label{eq:he-Friedmann}
H^2=\frac{8\pi G}{3}(\rho_{he}+\rho_{m})
\end{equation}
\begin{equation}\label{eq:he-Raychaudhuri}
\dot{H}=-4\pi G(\rho_{he}+\rho_m+p_{he}+p_m)
\end{equation}
where $\rho_m$ and $p_m$ are the energy density and pressure for
background matter. Ultimately, EOS parameter of Hessence is:
\begin{equation}\label{eq:he-state}
w_{he}=\frac{\dot{\phi^2}-\frac{Q^2}{a^6\phi^2}-2V(\phi)}{\dot{\phi^2}-\frac{Q^2}{a^6\phi^2}+2V(\phi)}
\end{equation}
And apparently when $\dot{\phi^2}\geqslant \frac{Q^2}{a^6\Psi^2}$,
Hessence EOS parameter $w_{he}\geqslant -1$；when
$\dot{\phi^2}\leqslant \frac{Q^2}{a^6\Psi^2}$, $w_{he}\leqslant -1$.
Via the evolution of $\phi$ field, EOS parameter $w$ crosses -1 when
$\dot{\phi^2}\geqslant \frac{Q^2}{a^6\Psi^2}$. Eqs
\ref{eq:Q}-\ref{eq:he-state} all depends on $Q^2$ straightly rather
than $Q$, which means $+Q$ and $-Q$
share the same dynamical behaviors, and leads to the assumption of existence of
Hessence DE and anti-Hessence DE.\\

Canonical and complex fields could easily generate the so-called
$Q-ball$ structure. Q-ball is a kind of non-topological soliton,
whose steadiness is related to the corresponding Q conserving
quantity. As to complex fields with canonical Lagrangian, such as
Spintessence, once the energy density fluctuates, the perturbation
would magnify nonlinearly at an exponential speed to form Q-balls
condensation. This a a big trouble for DE, because DE doesn't
condensate at the scales smaller than 100Mpc. Once come into being,
Q-balls act like matter and its energy density decay into other
particles at the speed proportional to $a^{-3}$. However, as for
Hessence DE, whose Lagrangian is noncanonical, the formation of
Q-balls are easily avoided. Although the Hessence Lagrangian never
occures before in quantum field theory, due to its steadiness
consequences, Hessence is treated highly as a reasonable complex
scalar field candidate for dark energy \cite{Hessence5}
\cite{Hessence6}.
\subsection{Hessence Reconstruction via Hubble Parameter}
Firstly,denote $M^2_{pl}\equiv (8\pi G)^{-1/2}$, and $M^2_{pl}$ is
the reduced Plank mass. Then (\ref{eq:he-Friedmann}),
(\ref{eq:he-Raychaudhuri}) could be rewritten as\cite{Hessence3}:
\begin{equation}
H^2=\frac{1}{3M^2_{pl}}(\rho_{he}+\rho_{m})
\end{equation}
\begin{equation}
\dot{H}=\frac{1}{2M^2_{pl}}(\rho_{he}+\rho_{m}+p_{he}+p_m)
\end{equation}
This is an interesting formulation. According to the two equations
above, one gets:
\begin{equation}\label{eq:he-new1}
V(\phi)=3M^2_{pl}H^2+M^2_{pl}\dot{H}-\frac{1}{2}\rho_m
\end{equation}
\begin{equation}\label{eq:he-new2}
\dot{\phi}^2-\frac{Q^2}{a^6\phi^2}=-2M^2_{pl}\dot{H}-\rho_m
\end{equation}
and the redshift being $z=a^{-1}-1$ (supposing $z_0=1$
 and subscript $0$ refers to today), then for arbitrary $f$,
\begin{equation}
\dot{f}=-(1+z)H\frac{df}{dz}
\end{equation}
and (\ref{eq:he-new1}), (\ref{eq:he-new2}) could be expressed as:
\begin{equation}\label{eq:he-new11}
V(z)=3M^2_{pl}H^2-M^2_{pl}(1+z)H\frac{dH}{dz}-\frac{1}{2}\rho_{m0}(1+z)^3
\end{equation}
\begin{equation}\label{eq:he-new22}
\left(\frac{d\phi}{dz}
\right)^2-\frac{Q^2}{\phi^2}(1+z)^4H^{-2}=2M^2_{pl}(1+z)^{-1}H_{-1}\frac{dH}{dz}-\rho_{m0}(1+z)H^{-2}
\end{equation}
now introduce such non-dimensional quantities as below:
\begin{equation}
\tilde{V}\equiv \frac{V}{M^2_{pl}H^2_0},\qquad
\tilde{\phi}\equiv\frac{\phi}{M_{pl}}
\end{equation}
\begin{equation}
\tilde{H}\equiv \frac{H}{H_0}, \qquad
\tilde{Q}\equiv{Q}{M^2_{pl}H_0}
\end{equation}
thus (\ref{eq:he-new11}), (\ref{eq:he-new22}) turn to:
\begin{equation}\label{eq:he-new111}
\tilde{V}(z)=3\tilde{H}^2-(1+z)\tilde{H}\frac{d\tilde{H}}{dz}-\frac{3}{2}\Omega_{m0}(1+z)^3
\end{equation}
\begin{equation}\label{eq:he-new222}
\left(\frac{d\tilde{\phi}}{dz}\right)-\tilde{Q}^2\tilde{\phi}^{-2}(1+z)^4\tilde{H}^{-2}=2(1+z)^{-1}\tilde{H}\frac{\tilde{H}}{dz}-3\Omega_{m0}(1+z)\tilde{H}^{-2}
\end{equation}
where $\Omega_{m0}=\rho_0/(3M^2_{pl}H^2_0)$ refers to the presentday
energy density of matter. Once $\tilde{H}(z)$ or $H(z)$ is known, we
could reconstruct $V(z)$ and $\phi(z)$ via (\ref{eq:he-new111}) and
(\ref{eq:he-new222}), and reconstruct $V(\phi)$ via $V(z)$ and
$\phi(z)$. From (\ref{eq:he-new2}) (\ref{eq:he-new111}) one gets the
reconstruction of Hessence EOS parameter:
\begin{equation}
w_{he}(z)\equiv\frac{p_{he}}{\rho_{he}}=\frac{-1+\frac{2}{3}(1+z)\frac{dln\tilde{H}}{dz}}{1-\Omega_{m0}\tilde{H}^{-2}(1+z)^3}
\end{equation}
Kinetic Energy could be induced from reconstruction of
(\ref{eq:he-new2}):
\begin{equation}
K\equiv \frac{\dot{\phi}^2}{2}-\frac{Q^2}{2a^6\phi^2}
\end{equation}
thus,
\begin{equation}
\tilde{K}(z)\equiv
\frac{K}{M^2_{pl}H^2_0}=(1+z)\tilde(H)\frac{d\tilde{H}}{dz}-\frac{3}{2}\Omega_{m0}(1+z)^3
\end{equation}
And this reconstruction depends on testable $H(z)$ rather than
certain models.
\subsection{Autonomous Equations for Hessence}
Firstly, to express the background matter via EOS of 正压 perfect
fluid:
\begin{equation}
p_m=w_m\rho_m\equiv(w-1)\rho_m
\end{equation}
where $0<\gamma<2$. Particularly, $\gamma=1$ and $\gamma=4/3$
represent the matter dominated state and radiation dominated state
respectively. Introduce the interaction term $C$ of Hessence and
matter at energy equilibrium, and when energy conservation conditon
$\rho_{tot}+3H(\rho_{tot}+p_{tot})=0$
holds\cite{Hessence2}\cite{Hessence4}, there's:
\begin{equation}
\dot{\rho}_{he}+3H(\rho_{he}+p_{he})=-C
\end{equation}
\begin{equation}
\dot{\rho}_m+3H(\rho_m+p_m)=C
\end{equation}
If there's no interaction between Hessence and background matter,
$C=0$, and interaction for $C\neq
0$.\\

To establish the differential autonomous equations, one takes the
following transformations, $\kappa^2=8\pi G$,$\kappa>0$, and
introduce the non-dimensional variables:
\begin{equation}
x\equiv\frac{\kappa \dot{\phi}}{\sqrt{6}H},\quad y\equiv\frac{\kappa
\sqrt{V}}{\sqrt{3}H},\quad
z\equiv\frac{\kappa\sqrt{\rho_m}}{\sqrt{3}H},\quad
u\equiv\frac{\sqrt{6}}{\kappa\phi},\quad
v\equiv\frac{\kappa}{\sqrt{6}H}\frac{Q}{a^3\phi}
\end{equation}
They forms the Hessence phase space, and now Hessence dynamical
problems can be treated as geometrical problems in Hessence phase
space. And according to basic Hessence equations one gets
\begin{equation}
x'=3x(x^2-v^2+\frac{\gamma}{2}z^2-1)-uv^2-\sqrt{\frac{3}{2}}y^2f-C_1
\end{equation}
\begin{equation}
y'=3y(x^2-v^2+\frac{\gamma}{2}z^2)+\sqrt{\frac{3}{2}xyf}
\end{equation}
\begin{equation}
z'=3z(x^2-v^2+\frac{\gamma}{2}z^2-\frac{\gamma}{2})+C_2
\end{equation}
\begin{equation}
u'=-xu^2
\end{equation}
\begin{equation}
v'=3v(x^2-v^2+\frac{\gamma}{2}z^2-1)-xuv
\end{equation}
where prime operator ($'$) refers to derivative to e-folding time
$N\equiv{lna}$, and
\begin{equation}
f\equiv\frac{1}{\kappa V}\frac{\partial V}{\partial \phi}
\end{equation}
\begin{equation}
C_1\equiv\frac{\kappa C}{\sqrt{6}H^2\dot{\phi}},\qquad
C_2\equiv\frac{\kappa C}{2\sqrt{3}H^2\sqrt{\rho_m}}=\frac{x}{z}C_1
\end{equation}
Thus Friedmann Equation reads:
\begin{equation}
x^2+y^2+z^2-v^2=1
\end{equation}
And the relative energy density for each component reads:
\begin{equation}
\Omega_{he}=\frac{\rho_{he}}{\rho_{crit}}=x^2+y^2-v^2,\qquad
\Omega_m=\frac{\rho_m}{\rho_{crit}}=z^2
\end{equation}
and the effective EOS parameters for Hessence and the whole cosmos
are:
\begin{equation}
w_{he}=\frac{p_{he}}{\rho_{he}}=\frac{x^2-v^2-y^2}{x^2-v^2+y^2}
\end{equation}
\begin{equation}
w_{eff}=\frac{p_{he}+p_m}{\rho_{he}+\rho_m}=x^2-v^2-y^2+(\gamma-1)z^2
\end{equation}
the critical points($\bar{x},\bar{y},\bar{z},\bar{u},\bar{v}$) is
gained via setting $\bar{x}'=\bar{y}'=\bar{z}'=\bar{u}'=\bar{v}=0$.
And this is the foundation for further analysis. Hessence has
perfect mathematical structure, so it is employed to show the normal
process of dynamical analysis here.\\

It is shown in \cite{Hessence2} that, as for exponential potential
and power potential,  all late attractor solutions meet
$\omega_{he}\geq -1$ and $\omega_{eff}\geq -1$ with 4-form
interaction term $C$. Nevertheless, Phantom style solutions with
$\omega_{he}<-1$ or $\omega_{eff}<-1$ is not stable, and will evolve
into Quintessence attractor with $\omega_{he}\geq -1$ or
$\omega_{eff}\geq -1$, or de Sitter attractor with $\omega_{he}= -1$
or $\omega_{eff}= -1$. So Big Rip is avoided in Hessence
cosmology.\\

In \cite{Hessence4}, it is shown that in 4 dimensional phase space,
the existence of stable late attractor solutions is independent of
the forms of $V(\phi)$ and C, but the stability of the solutions
depends on the second order derivative of Hessence potential
$V(\phi)$.

\chapter{k-essence \& Chaplygin gas }
\section{k-essence Mechanism}
The Scalar field dark energy models employ slowly rolling potentials
to drive the accelerating expansion. Yet, such a cosmos scenario can
be gained via modifying the kinetic energy term. In 1999,
Armendariz-Picon realized that the kinetic energy term could lead to
cosmic inflation in the baby era that is governed by high energy
mechanism, and introduced k-inflation. Afterwards Chiba utilized the
same mechanism to describe dark energy that drive the accelerating
expansion, and Armendariz-Picon improved this work into k-essence.
Fork-essence, high order modifications appears at the kinetic terms
to drive the acceleration. And Bohn-Infeld field is one form of
k-essence. If the form of pressure is properly selected, the
fine-tuning problem and coincidence problem can also be solved\cite{K-ess1} \cite{K-ess2} \cite{k-ess3}.\\

k-essence has noncanonical kinetic energy. As for a scalar field
with $\phi$ and kinetic energy $X\equiv -(1/2)(\nabla \phi)^2$, the
action with the most generalized form is:
\begin{equation}
S=\int d^4 x\sqrt{-g} p(\phi, X)
\end{equation}
where Lagrangian density is the same with pressure density $p(\phi,
X)$. and the action is also used in generalized Quintessence models.\\

Usually, Lagrangian density of k-essence obeys the following
constraints:
\begin{equation}\label{eq:k-lag}
p(\phi,X)=f(\phi)\hat{p}(X)
\end{equation}
This constraints originates from superstring theory, in which low
energy effect will leads to higher order derivative terms from $a'$
and loop modifications (the relation of $a'$ and string length
scalar $\lambda_s$ is $a'=\lambda_s /2\pi$). The four dimensional
effective string action is:
\begin{equation}
S=\int
d^4x\sqrt{-\tilde{g}}\{B_g(\phi)\tilde{R}+B^{(0)}_\phi(\phi)(\tilde{\nabla}\phi)^2-a'[c_1^{(1)}B_\phi^{(1)}(\phi)(\tilde{\nabla}\phi)^4+\ldots]+O(a'^2)
\}
\end{equation}
where $k^2=8\pi G=1$. Here $\phi$ is Dilaton field, which controls
the string coupling strength $g^2_s$ via $g^2_s=e^\phi$. For weak
couplings $e^\phi<<1$, the coupling equation is $B_g\approx
B_\phi^{(0)}\approx B_\phi^{(1)}\approx e^{-\phi}$. If the coupling
is one order, it will take a more complicated form. Via conformal
transformation $g_{\mu\nu}=B_g(\phi)\tilde{g}_{\mu\nu}$, the string
action will transforms into Einsteinian action form:
\begin{equation}
S_E=\int d^4x\sqrt{-g}[\frac{1}{2}R+K(\phi)X+L(\phi)X^2+\ldots]
\end{equation}
with the component
\begin{equation}
K(\phi)=\frac{3}{2}\left(\frac{1}{B}\frac{dB_g}{d\phi}
\right)^2-\frac{B^{(0)}}{B_g}
\end{equation}
\begin{equation}
L(\phi)=2c_1^{(1)}a'B_\phi^{(1)}(\phi)
\end{equation}
Thus we could get the Lagrangian with noncanonical kinetic energy
term:
\begin{equation}
p(\phi,X)=K(\phi)X+L(\phi)X^2
\end{equation}
and redefine the field $\phi_{new}$ as:
\begin{equation}
\phi_{new}=\int ^{\phi_{old}}d\phi\sqrt{\frac{L}{|K|}}
\end{equation}
Thus Lagrangian transforms into:
\begin{equation}
p(\phi,X)=f(\phi)(-X+X^2)
\end{equation}
where $\phi=\phi_{new}$,$X\equiv X_{new}=(L/|K|)X_{old}$,and
$f(\phi)=K^2(\phi_{old})/L(\phi_{old})$. This is the k-essence model
induced from (\ref{eq:k-lag})via $\hat{p}(X)=-X+X^2$. Thus the
pressure is also known, thus the energy density of the scalar field
$\phi$ is:
\begin{equation}
\rho=2X\frac{\partial p}{\partial X}-p=f(\phi)(-X+3X^2)
\end{equation}
With energy density and pressure, one immediately gets EOS parameter
for $\phi$:
\begin{equation}\label{eq:k-state}
w_\phi=\frac{p}{\rho}=\frac{1-X}{1-3X}
\end{equation}
and as we can see, for certain $X$, $w_\phi$ is also a constant. For
example, when $X=1/2$ we could get a cosmological constant style EOS
parameter $w_\phi=-1$; to get an accelerating expansion $w<-1/3$,one
gets $X<2/3$.\\

According to energy density $\rho$, one gets the continuity equation
(\ref{eq:ener-monen-conserve}). When matter and radiation is
dominant and EOS parameter for background fluid is $w_m$,
Eq(\ref{eq:Friedmann2}) gives rise to the Hubble parameter:
\begin{equation}
H=\frac{2}{3(1+w_m)(t-t_0)}
\end{equation}
and the energy density being
\begin{equation}
\dot{\rho}=-\frac{2(1+w_\phi)}{(1+w_m)(t-t_0)}\rho
\end{equation}
For certain $X$, or certain $w_\phi$, $f(\phi)$ is constraint for
\begin{equation}
f(\phi)\propto(\phi-\phi_0)^{-\alpha},\quad
\alpha=\frac{2(1+w_\phi)}{1+w_m}
\end{equation}

when $w_\phi=w_m$, $f(\phi)\propto(\phi-\phi_0)^{-2}$ for radiation
dominated or matter dominated era, and this agrees with the scaling
solutions. Thus, as for dark energy, one must take fine tuning of
$f(\phi)$ to satisfy the presentday cosmic density. And one should
note that the density of the scalar field must be significantly less
than that of the background fluid, i.e. $\rho\ll\rho_m$, which
contradicts with a DE dominated universe. For example,
$f(\phi)\propto(\phi-\phi_0)^2$, except for a late scaling solution,
we has another solution of accelerating expansion. Actually this is
the divide of accelerating expansion with decelerating expansion. \\

From (\ref{eq:k-state}) we could know that, the kinetic term $x$
plays the key role to influent the EOS. If $1/2<X<2/3$ holds, the
scalar field $\phi$ will acts as dark energy with $0\leqslant \alpha
\leqslant 2 $. Armendariz-Picon et al have made more generalized
analysis with $\hat{p}(X)$ to avoid the fine-tuning problem.

\section{Chaplygin gas model}
In the discussions with Quintessence, the background perfect fluid
always has a EOS of $p=w \rho$, and change different potentials and
Lagrangian. What if employing some other form of EOS ? It is very
coincident that, when three superstring experts were studying the
stability of black holes in brane world, they found they have to add
a background matter called Chaplygin gas, to ensure the stability.
Chaplygin gas is a perfect fluid with negative pressure in
superstring theory. They proceeded to calculate the cosmological
consequence for a FRW universe comprised of matter, CDM and
chaplygin gas, and found that a flat universe with accelerating
expansion is gained, and chaplygin gas not only acts as dark energy,
but strongly indicates a unified description of dark energy and dark
matter. Hence, Chaplygin gas is introduced as a DE candidate in a quite natural manner.\\
\subsection{Special Chaplygin gas}
EOS of Chaplygin gas is (all components depend on comoving
coordinates):
\begin{equation}\label{eq:chaply-state}
p_{CG}=-\frac{A}{\rho_{CG}}
\end{equation}
where $\rho_{CG}>0$, $p_{CG}>0$, $A$ is a positive constant.
Chaplygin gas obeys the energy-momenta equation
(\ref{eq:ener-monen-conserve}), (\ref{eq:thermal1}). Insert the EOS
of Chaplygin gas into (\ref{eq:thermal1}), and the integral induces:
\begin{equation}\label{eq:chaply-old-2}
\rho_{CG}=\sqrt{A+\frac{B}{a^6}}
\end{equation}
Where B is an integral constant, and is selected to be positive in
the following discussions; $a$ is the scalar factor.
Eq(\ref{eq:chaply-old-2}) show the behaviors:
\begin{equation}
a\ll(\frac{B}{A})^{\frac{1}{6}}\longrightarrow\rho_{GC}\sim
\frac{\sqrt{B}}{a^3}
\end{equation}
\begin{equation}\label{eq:chaply1}
a\gg(\frac{B}{A})^{\frac{1}{6}}\longrightarrow\rho_{CG}\sim-p_{CG}\sim
\sqrt{A}
\end{equation}
And we can easily draw the conclusions that, at early epoches, $a$
is small enough to ensure $a\ll(\frac{B}{A})^{\frac{1}{6}}$, and
density of Chaplygin gas $\rho_{CG}\sim a^{-3}$, which is like
non-relativistic matters; lately $a$ is large enough to ensure
$a\gg\frac{B}{A})^{\frac{1}{6}}$,thus $\rho_{CG}\sim-p_{CG}\sim
\sqrt{A}$ and acts like the cosmological constant, which has
negative pressure to drive the accelerating expansion, and the time
needed to arrive at this state is:
\begin{equation}
t=\frac{1}{6\sqrt[4]{A}}\left(\ln\frac{\sqrt[4]{A+\frac{B}{a^6}}+\sqrt[4]{A}}{\sqrt[4]{A+\frac{B}{a^6}}-\sqrt[4]{A}}-2\arctan
\sqrt[4]{1+\frac{B}{Aa^6}} \right)
\end{equation}
Because Chaplygin gas acts like dark matter and dark energy at
different epoches, it indicates a unification of CDM and DE.\\

A more detailed perturbative form of (\ref{eq:chaply1}) is:
\begin{equation}\label{eq:chaply2}
\rho\sim\sqrt{A}+\sqrt{\frac{B}{4A}}a^6
\end{equation}
\begin{equation}\label{eq:chaply3}
p\sim-\sqrt{A}+\sqrt{\frac{B}{4A}}a^6
\end{equation}
(\ref{eq:chaply2}), (\ref{eq:chaply3}) show that the expanding
universe is made up of two components, the major part acts like the
cosmological constant with $p=-\rho$; the minor part acts as
$p=\rho$, whose energy density decrease sharply with the
expansion.\\

The cosmological consequences of Chaplygin gas could be treated as a
homogeneous scalar field $\phi(t)$ togther with a potential
$V(\phi)$. Consider the Lagrangian:
\begin{equation}
\mathcal{L}_{GC}(\phi)=\frac{1}{2}\dot{\phi}^2-V(\phi)
\end{equation}
and set the density of the field equal to that of Chaplygin gas:
\begin{equation}
\rho_\phi=\frac{1}{2}\dot{\phi}^2+V(\phi)=\sqrt{A+\frac{B}{a^6}}
\end{equation}
the pressure is identical to Lagrangian:
\begin{equation}
p_\phi=\frac{1}{2}\dot{\phi}^2-V(\phi)=-\frac{A}{\sqrt{A+\frac{B}{a^6}}}
\end{equation}
After some algebra, the reasonable potential is:
\begin{equation}
V(\phi)=\frac{1}{2}\sqrt{A}\left(\cosh 3\phi+\frac{1}{\cosh 3\phi}
\right)
\end{equation}
The potential is independent of B, and it reflects the properties
EOS of Chaplygin gas (\ref{eq:chaply-state}) only.\\

There are several approaches to EOS of Chaplygin gas, including:\\

(1)Action and Potential of Quintessence
\begin{equation}
V(\phi)=\frac{\sqrt{a}}{2}(\cosh 3\phi+\frac{1}{\cosh 3\phi})
\end{equation}
and in return, with the factors A, B and scale factor $a$ in
Chaplygin gas, one can re-express the kinetic energy and potential
in Quintessence:
\begin{equation}
\dot{\phi}^2=\frac{B}{a^6\sqrt{A+B/a^6}}
\end{equation}
and
\begin{equation}
V(\phi)=\frac{2a^6(A+B/a^6)-B}{2a^6\sqrt{A+B/a^6}}
\end{equation}
(2)Born-Infeld Lagrangian
\begin{equation}
\mathcal {L}=-V_0\sqrt{1-\partial_\mu\partial^\mu}
\end{equation}
has the induction that:
\begin{equation}
\rho=\frac{V_0}{\sqrt{1-\partial_\mu\partial^\mu}}
\end{equation}
\begin{equation}
p=-V_0\sqrt{1-\partial_\mu\partial^\mu}
\end{equation}
which end up with Chaplygin gas EOS.

\subsection{Generalized Chaplygin gas}
Firstly to generalize EOS of Chaplygin gas(GCG) into:
\begin{equation}\label{eq:chaply-state2}
p=-\frac{A}{\rho^\beta}
\end{equation}
where $0<\beta\leq 1$.(further consraints sre mage in [24] to get
$0\leqslant \beta \leqslant 0.5 $. The following task is to find out
the scalar field corresponding to (\ref{eq:chaply-state2})\\

For a complex field with non-zero mass, supposing:
\begin{equation}
\Phi=\left(\frac{\phi}{\sqrt{2}}m \right)\exp \left(-im\theta
\right)
\end{equation}
with Lagrangian of:
\begin{equation}
\mathcal{L}_{GCG}=g^{\mu\nu}\left(\phi^2\theta_{,\mu}\theta_{,\nu}
\right)-V(\phi^2/2)
\end{equation}
where $,\mu$ refers to variance with spatial coordinates $x^\mu$ .
The condition $\phi_{,\mu}\ll m\phi$ is used to ensure the
fluctuations of energy density in spacetime, and together with the
generalized EOS (\ref{eq:chaply-state2}) and Thomas-Fermi, to
rewrite its Lagrangian:
\begin{equation}\label{eq:chaply-lagrangian}
\mathcal{L}_{TF}=\frac{\phi^2}{2}g^{\mu\nu}\theta_{,\mu}\theta_{,\nu}
\end{equation}
This acts as first principle for generalized Chaplygin gas cosmology.\\

From(\ref{eq:chaply-state2}), one has a generalized
(\ref{eq:chaply-old-2}):
\begin{equation}
\rho=\left(A+\frac{B}{a^{3(1+\beta)}}  \right)^{\frac{1}{1+\beta}}
\end{equation}
Energy density for generalized Chaplygin gas being:
\begin{equation}
\rho_{GCG}=(A+\frac{B}{a^{3(1+\beta)}})^{\frac{1}{1+a}}
\end{equation}
the effective EOS is:
\begin{equation}
\omega(a)=-\frac{|\omega_0|}{|\omega_0|+\frac{1-|\omega_0|}{a^{3(1+\beta)}}}
\end{equation}
and the perturbations (\ref{eq:chaply2}), (\ref{eq:chaply3}) turn to
be:
\begin{equation}
\rho\sim A^{\frac{1}{1+\beta}}+\left(\frac{1}{1+\beta}
\right)\frac{B}{A^{\frac{\beta}{1+\beta}}}a^{-3(1+\beta)}
\end{equation}
\begin{equation}
p\sim -A^{\frac{1}{1+\beta}}+\left(\frac{1}{1+\beta}
\right)\frac{B}{A^{\frac{\beta}{1+\beta}}}a^{-3(1+\beta)}
\end{equation}
Thus, as for the generalized Chaplygin gas, the expanding cosmos is
also made up of two components. The major part is dark energy with
$p=\rho$ and $A^{\frac{1}{1+\beta}}$. The minor part is of EOS
$p=\beta \rho$ now, and the potential for generalized Chaplygin gas
with Lagrangian (\ref{eq:chaply-lagrangian}) is:
\begin{equation}
V=\frac{1}{2}\left(\Psi^{2/\beta}+\frac{A}{\Psi^2} \right)
\end{equation}
where $\psi\equiv B^{(1-\beta/1+\beta)}a^{3(1-\beta)}\phi^2$.\\

There's some other generalized form, for example
\begin{equation}
p=-\frac{A(a)}{\rho}
\end{equation}
where $A(a)$ depends on the scale factor $a$, and is set positive. A
modified or generalized version can be more flexible and can supply
better cosmic scenario.
\subsection{Generalized Chaplygin gas \& Hessence}
If the field is properly selected, the late behaviors of Hessence
complex scalar field could be expressed via generalized Chaplygin
gas. Firstly, Hessence should obey EOS (\ref{eq:chaply-state2}):
\begin{equation}
p_{he}=-\frac{A}{\rho^\beta_{he}}
\end{equation}
Via EOS of Hessence (\ref{eq:he-state}):
\begin{equation}
2V(\phi)=\rho_{he}-p_{he}=\rho_{he}+\frac{A}{\rho^\beta_{he}}
\end{equation}
\begin{equation}
\dot{\phi}-\frac{Q^2}{a^6\phi^2}=\rho_{he}+p_{he}=\rho_{he}-\frac{A}{\rho^\beta_{he}}
\end{equation}
Friedmann Equation(\ref{eq:Friedmann}) for  Hessence dominated
\begin{equation}
\left(\frac{\dot{a}}{a}  \right)^2=H^2=\frac{8\pi G}{3}\rho_{he}
\end{equation}
If the above condition holds,
\begin{equation}
\dot{\phi}^2\ll \frac{Q^2}{a^6\phi^2}, \qquad \beta=1
\end{equation}
Analyzable solutions are achieved:
\begin{equation}
\rho_{he}=\frac{B\phi^2}{Q^2},\qquad
a=\left(\frac{BQ^4}{AQ^4-B^2\phi^4} \right)^{1/6}
\end{equation}
\begin{equation}
\dot{\phi}^2=\frac{6\pi GBQ^2}{a^12(A-B/a^6)}
\end{equation}

\subsection{Chaplygin gas \& Superstring QFT}
The origin of Chaplygin gas cosmology has shown its connections with
superstring quantum field theory. Chaplygin gas is the
non-relativistic limit of Born-Infeld mechanism. For excellent
details, see Jackiw\cite{chaply3}.

\chapter{Hyperbolic Cosecant Cardassian Cosmology}
\section{Introduction}
In 2002, K. Freese and M.Lewis from Michigan University came up with
a new mechanism for dark energy, where dark energy doesn't exist at
all\cite{Card1}! They modified Friedmann equation, or the
gravitational field equation, which is of fundamental significance
in cosmological dynamics, to  establish a scenario where the
universe is flat, matter dominated, and has accelerating expansion
recently. They treated matter and radiation together (denoted as a
matter term), and introduced a new term, Cardassian energy density,
which depends quantitatively only on the density of the matter term,
to the right side of Friedmann equation (the original discussion
takes power function style Cardassian term as an example). When
dated back to the baby era, the Friedmann equation would return to
its normal form with no appearance of Cardassian term, and the
process such as nuclearsynthesis and galaxies formation are not
disturbed. Cardassian term turns dominant only recently, with
redshift $z\sim O(1)$. In Cardassian cosmos, the critical density is
remarkably less than the normal value, and can be exactly
$\Omega_{crit}=\Omega_m$ via tuning parameters. So, in Cardassian
world, matter itself determines the flatness of the universe, and
Cardassian term takes over to drive an accelerating expansion
recently. There is no room for dark energy any more. Cardassian
cosmology has survived several observational test, such as the age
of the universe, CMB, etc. Initially, the origin of Cardassian term
is attributed to the natural consequence when we insert the
observational cosmos into a world with higher
dimension.\\

But the work of K. Freese and M. Lewis didn't raised much attention
originally. So, 7 months later Freese expressed the idea again and
generalize the density term in Friedmann equation to a general form
$g(\rho)$ by the way \cite{Card2}. Shortly, Gondolo and Freese put
forward the relativistic fluid interpretation of Cardassian dynamics
\cite{Card3}(denoted as GF fluid hereafter). Cardassian cosmology is
raised to the hight of theory from phenomenology, and the dynamical
mechanism is analysed comprehensively. This work is a milestone in
the study of Cardassian cosmology, which lay the foundation for
future research on this subject, and up to now, nearly all the
discussions with Cardassian are based on GF fluid scenario. To
ensure the positiveness of sound speed, they analysed a new kind of
Cardassian term, the polytropic and a better modified polytropic
style. They also come up with a proposal that the negative pressure
arise from the interactions between dark matter. From this thesis
on,
Cardassian cosmology wins large attention and hot discussions.\\

In \cite{Card4} it is shown that power style Cardassian violates the
weak energy conditions of gravitation on small scales. Although
suffering this problem, power style Cardassian, due to its simplest
form, is still widely employed when new ideas are introduced. Two
other important work attribute to \cite{card5}(M Szydlowski and W
Czaja) and \cite{card6}(R Lazkoz, G Leon). The former provides an
excellent example for autonomous dynamics and numerical analysis for
Cardassian. The latter divides the Cardassian discussion of
autonomous dynamical system into high energy limit and low energy
limit two categories for the first time. The typical work to
generalize Cardassian model includes the introduction of exponential
Cardassian term\cite{card8}; Statefinder diagnose for modified
polytropic Cardassian\cite{card9}. Generally, computations with
Cardassian cosmology require the background to be homogeneous and
isotropic, and thesis \cite{card10} makes an attempt to cancell the
restriction of homogeneity. Meantime, numerical analysis and
observational constraints for Cardassian take step \cite{Carde1}
\cite{Carde2} \cite{Carde3} \cite{Carde4} \cite{carde5}
\cite{carde6} \cite{carde7} \cite{carde8} \cite{carde9}
\cite{carde10} \cite{carde11} \cite{carde12} \cite{carde13}
\cite{carde14}, mainly with CMB, the cosmic age, gravitational
lensing
and large scale exploring.\\

Freese and her cooperators have written several reviews on
Cardassian cosmology, which does good for a glance of Cardassian,
see \cite{Card2} \cite{card7-1} \cite{card7-2}. And in this chapter,
we will do a job similar to \cite{card8}, to introduce a new model
via the introduction of hyperbolic cosecant Cardassian term.
\section{Fundamentals of Cardassian Cosmology}
As expressed above, the following analysis is based on GF fluid
scenario. In the original work, matter and radiation are treated
together for a matter energy density term and the independent
variable of Cardassian term. The following analysis will change into
two cases. This treatment is reasonable, because Cardassian term is
effective only recently ($z\sim O(1)$) when matter is far more
enriching than radiation, and radiation is considerable and makes a
difference only in the early era, when the Cardassian modification
is neglected. But for a clearer scenario or as a generalization, to
treat matter and radiation separately is also a good choice. So in
this text two cases are discussed, namely treating matter and
radiation together as a matter term and treating them separately.
\subsection{Fundamentals for Implicit Cardassian term}
In Cardassian cosmology, the total effective energy density
is\cite{Card1}\cite{Card2}\cite{Card3}:
\begin{equation}\label{eq:card1}
\rho'=g(\rho)
\end{equation}
where $\rho=\rho_m+\rho_\gamma$ reflects the summary of energy
density of matter  $\rho_m$ and energy density of radiation
$\rho_\gamma$ , which are testable parameters. From the first law
and second law of thermaldynamics, we have $Td(sV)=d(\rho V)+pdV$;
for adiabatic expanding $d(sV)=0$; conservation of particles' number
$d(\rho_m V)=0$; conservation of radiation $d(\rho_\gamma
V^{\frac{4}{3}})=0$. Insert these relations into (\ref{eq:card1})
and we could get the total pressure:
\begin{equation}\label{eq:card2}
p'=-\rho'+\rho_m\frac{\partial \rho'}{\partial
\rho_m}+\frac{4}{3}\rho_\gamma\frac{\partial \rho'}{\partial
\rho_\gamma}
\end{equation}
In Cardassian cosmos, Friedmann equation reads:
\begin{equation}
H^2=\frac{8\pi G}{3}g(\rho)
\end{equation}
and Raychaudhuri equation reads:
\begin{equation}
\dot{H}=-4\pi G(\rho_m\frac{\partial \rho'}{\partial
\rho_m}+\frac{4}{3}\rho_\gamma\frac{\partial \rho'}{\partial
\rho_\gamma})
\end{equation}
the energy-momentum conservation equation:
\begin{equation}
\dot{\rho}'=-3H(\rho'+p')
\end{equation}
The above equations are the fundamental equations for Cardassian
cosmology.
\subsection{Fundamentals for Explicit Cardassian term}
Firstly dividing $\rho'=g(\rho)$ into two terms:
\begin{equation}
\rho'=\rho+f(\rho)
\end{equation}
The first term $\rho$ is the matter and radiation term, and
$f(\rho)\equiv g(\rho)-\rho$ is the explicit Cardassian term, which
is the modification term of Friedmann equation. When matter and
radiation are treated separately:
\begin{equation}
\rho'=\rho_m+\rho_\gamma+f(\rho_m,\rho_\gamma)
\end{equation}
Thus from (\ref{eq:card2}) one gets the total energy:
\begin{equation}\label{eq:card7}
p'=\frac{\rho_\gamma}{3}+\rho_m\frac{\partial f}{\partial
\rho_m}+\frac{4}{3}\rho_\gamma\frac{\partial f}{\partial
\rho_\gamma}-f
\end{equation}
and from this equation, we can easily see that only radiation and
Cardassian term contribute to the total pressure, or EOS parameter
for matter is $w=0$. This conclusion will be quoted in next chapter
for virilization. Friedmann equation and Raychaudhuri equation are
\begin{equation}
H^2=\frac{8\pi G}{3}[\rho_m+\rho_\gamma+f(\rho_m,\rho_\gamma)]
\end{equation}
and
\begin{equation}
\dot{H}=-4\pi G(\rho_m+\frac{4}{3}\rho_\gamma+\rho_m\frac{\partial
f}{\partial \rho_m}+\frac{4}{3}\rho_\gamma\frac{\partial f}{\partial
\rho_\gamma})
\end{equation}
And energy conservation is equal to the conservation of particles'
number and conservation of radiation:
\begin{equation}
\dot{\rho}_m=-3H\rho_m
\end{equation}
\begin{equation}
\dot{\rho}_\gamma=-4H\rho_\gamma
\end{equation}
\subsection{Phase Space, Autonomous Equations for Separate Matter \& Radiation}
Based on Friedmann equation, we define
\begin{equation}
1=\frac{8\pi G}{3H^2}[\rho_m+\rho_\gamma+f(\rho_m,\rho_\gamma)]
\end{equation}
$$
\equiv \Omega_m+\Omega_\gamma+\Omega_{card}
$$
where $\Omega_m$, $\Omega_\gamma$, $\Omega_{card}$ are the relative
density parameter of matter, radiation, Cardassian term
respectively. Now we introduce the non-dimensional parameters (with
$\kappa^2=8\pi G$):
\begin{equation}
x=\frac{\kappa \sqrt{\rho_m}}{\sqrt{3}H},\qquad y=\frac{\kappa
\sqrt{\rho_\gamma}}{\sqrt{3}H},\qquad z=\frac{\kappa
\sqrt{f}}{\sqrt{3}H},\qquad N=\ln a
\end{equation}
where N is e-folding time. Apparently $N$ keeps the same pace with
the cosmological time in our flat, expanding FRW universe. Those
newly introduced variables form the 4 dimensional phase space, and
when $x$,$y$,$z$, $N$ are inserted into the basic dynamical
equations, they will turn to the geometric curves in the phase space
by:
\begin{subequations}\label{eq:card-1}
\begin{equation}
\frac{dx}{dN}=-\frac{3}{2}x+\frac{3}{2}x^3+2xy^2+\frac{3}{2}x^3\frac{\partial
f}{\partial \rho_m}+2xy^2\frac{\partial f}{\partial \rho_\gamma}
\end{equation}
\begin{equation}
\frac{dy}{dN}=-2y+\frac{3}{2}yx^2+2y^3+\frac{3}{2}yx^2\frac{\partial
f}{\partial \rho_m}+2y^3\frac{\partial f}{\partial \rho_\gamma}
\end{equation}
\begin{equation}
\frac{dz}{dN}=\frac{3}{2}x^2\left(z-\frac{1}{z}
\right)\frac{\partial f}{\partial \rho_m}+2y^2\left(z-\frac{1}{z}
\right)\frac{\partial f}{\partial \rho_\gamma}+\frac{3}{2}x^2z+2y^2z
\end{equation}
\end{subequations}
These curves are called the differential autonomous equations of the
system. For further discussion, we treat the cosmos as flat (the
confirmed result by CMB etc), then there will be the restriction
$x^2+y^2+z^2=1$, which simplify the equations (\ref{eq:card-1})
above into:
\begin{subequations}\label{eq:card-2}
\begin{equation}
\frac{dx}{dN}=-\frac{3}{2}x+\frac{3}{2}x^3+2xy^2+\frac{3}{2}x^3\frac{\partial
f}{\partial \rho_m}+2xy^2\frac{\partial f}{\partial \rho_\gamma}
\end{equation}
\begin{equation}
\frac{dy}{dN}=-2y+\frac{3}{2}yx^2+2y^3+\frac{3}{2}yx^2\frac{\partial
f}{\partial \rho_m}+2y^3\frac{\partial f}{\partial \rho_\gamma}
\end{equation}
\end{subequations}
Particularly, if the Cardassian term has the symmetry
$f(\rho_m,\rho_\gamma)=f(\rho_m+\rho_\gamma)$, the equations
(\ref{eq:card-2}) can be further reduced into:
\begin{subequations}\label{eq:card9}
\begin{equation}
\frac{dx}{dN}=-\frac{3}{2}x+\left(\frac{3}{2}x^3+2xy^2
\right)\left(1+\frac{\partial f}{\partial \rho_m} \right)
\end{equation}
\begin{equation}
\frac{dy}{dN}=-2y+\left(\frac{3}{2}x^2y+2y^3
\right)\left(1+\frac{\partial f}{\partial \rho_m} \right)
\end{equation}
\end{subequations}
This case is important, because the hyperbolic cosecant Cardassian
term, as we will put forward later, just has such a lovely
character.\\

So, once $f(\rho_m,\rho_\gamma)$ is known, we will master the
dynamics of the Cardassian universe, and the proximation behaviors
of the corresponding autonomous equations in phase space.
\subsection{Unitary Treatment of Matter \& Radiation}
If matter and radiation are treated together as a matter term, or
only take matter into account for recent universe, the energy total
density will be
\begin{equation}\label{eq:card3}
\rho'=\rho_m+f(\rho_m)
\end{equation}
From(\ref{eq:card2}) one gets:
\begin{equation}
p'=\rho_m\frac{\partial f}{\partial \rho_m}-f
\end{equation}
Friedmann equation and Raychaudhuri are separately:
\begin{equation}
H^2=\frac{8\pi G}{3}[\rho_m+f(\rho_m)]
\end{equation}
and
\begin{equation}
\dot{H}=-4\pi G\rho_m(1+\frac{\partial f}{\partial \rho_m})
\end{equation}
Now the energy conservation equation $\dot{\rho'}=-3H(\rho'+p')$ is
equal to the conservation of particles' number
$\dot{\rho_m}=-3H\rho_m $. When matter and radiation is treated
together, there's convenient form for the speed of sound, which
should be positive:
\begin{equation}\label{eq:card10}
c_s^2=\frac{\partial p'}{\partial
\rho'}=\rho_m\frac{\frac{\partial^2\rho'}{\partial
\rho^2_m}}{\frac{\partial \rho'}{\partial \rho_m}}
\end{equation}
Based on Friedmann equation, we define
\begin{equation}
1=\frac{8\pi G}{3H^2}[\rho_m+f(\rho_m)]
\end{equation}
$$
\equiv \Omega_m+\Omega_{card}
$$
and introduce the new variables ($\kappa^2=8\pi G$):
\begin{equation}
x=\frac{\kappa\sqrt{\rho_m}}{\sqrt{3}H},\qquad
y=\frac{\kappa\sqrt{f}}{\sqrt{3}H}
\end{equation}
Together with e-folding time this leads to a 3-dimensional phase
space. Insert $x$, $y$ into the equations above to get:
\begin{subequations}\label{eq:card5}
\begin{equation}
\frac{dx}{dN}=-\frac{3}{2}x+\frac{3}{2}\left(1+\frac{\partial
f}{\partial \rho_m} \right)
\end{equation}
\begin{equation}
\frac{dy}{dN}=-\frac{3}{2}\frac{x^2}{y}\frac{\partial f}{\partial
\rho_m}+\frac{3}{2}x^2y\left(1+\frac{\partial f}{\partial \rho_m}
\right)
\end{equation}
\end{subequations}
with constraint $x^2+y^2=1$ for flat FRW space, the number of
independent equations reduces to one:
\begin{equation}
\frac{dx}{dN}=-\frac{3}{2}x+\frac{3}{2}\left(1+\frac{\partial
f}{\partial \rho_m} \right)
\end{equation}
Also, we know the universe if we know $f(\rho_m)$. And for simple
Cardassian terms, we can even get the exact and analyzable solution.
\subsection{Constraints for Cardassian \& Hyperbolic cosecant Cardassian}
Based on the general discussion above, we can make detailed analysis
with certain form of Cardassian term. Yet the Cardassian term should
not be arbitrary, as has been shown in the above algebra, and a
potential candidate must obey at least three
constraints \cite{card8}:\\

(1) As for the baby universe with $\rho_{crit,0}\ll \rho$, total
energy density $g(\rho)$ must reduces to $\rho$; or Cardassian term
$f(\rho)$ vanishes. \\

(2)Cardassian term is dominant recently with redshift $z\sim O(1)$,
when $g(\rho)$ differs from $\rho$ significantly, or $f(\rho)$
differs from null significantly.\\

(3)The speed of sound $c_s^2>0$ must be positive.\\

The constraints is in fact powerful. Only those that satisfies the
first condition is worth further algebra during which the second and
third condition proceeds to be tested. Three set of reasonable
Cardassian terms have been introduced before:\\

(1)Power-style Cardassian\cite{Card1}
\begin{equation}
\rho'=\rho+b\rho^n
\end{equation}
where $b$ and $n$
is tuning parameters, and $n<2/3$.\\
(2)Polytropic and Modified Cardassian\cite{Card3}
\begin{subequations}
\begin{equation}
\rho'=\rho_{card}\left[1+\left(\frac{\rho_m}{\rho_{card}}  \right)^q
\right]^{\frac{1}{q}}
\end{equation}
\begin{equation}
\rho'=\rho_{card}\left[1+\left(\frac{\rho_{card}}{\rho_m}
\right)^{qv} \right]^{\frac{1}{q}}
\end{equation}
\end{subequations}
where $q$ and $v$ is tuning parameters; $\rho_{card}$ a constant density value for reference.\\
(3)Exponential and Modified Cardassian\cite{card8}
\begin{subequations}
\begin{equation}
\rho'=\rho \exp \left(\frac{\rho_{card}}{\rho}  \right)^n
\end{equation}
\begin{equation}
\rho'=(\rho+\rho_{card})\exp\left(\frac{q\rho_{card}}{\rho+\rho_{card}}
\right)^n
\end{equation}
\end{subequations}
where $q$ and $n$ is tuning parameters.\\

Next we will discuss a new Cardassian term, namely hyperbolic
cosecant Cardassian term:
\begin{equation}
\rho'=\rho\left(1+b\cdot csch\left(\frac{\rho}{\rho_{card}}
\right)^n\right)
\end{equation}
where $b,n$ are tuning parameters and
\begin{equation}
f(\rho)=\rho b\cdot csch\left(\frac{\rho}{\rho_{card}} \right)^n
\end{equation}

Hyperbolic cosecant function differs very much from double
exponential potential function. Before further discussion, it's a
convenience to review the properties of hyperbolic cosecant
function.
\section{Fundamentals of Hyperbolic Cosecant Function} Hyperbolic sine $sh(x)$ and hyperbolic cosine $ch(x)$
are the two basic functions which are orthogonal to each other in
the hyperbolic function space, and are defined as:
\begin{equation}
sh(x)=\frac{e^x-e^{-x}}{2}   \qquad ch(x)=\frac{e^x+e^{-x}}{2}
\end{equation}
And definition of hyperbolic cosecant function is:
\begin{equation}
csch(x)=\frac{1}{sh(x)}=\frac{2}{e^x-e^{-x}}
\end{equation}
Its one order derivative is:
\begin{equation}
\frac{dcsch(x)}{dx}=-csch(x)cth(x)
\end{equation}
Where $cth(x)$ is the hyperbolic cotangent function:
\begin{equation}
cth(x)=\frac{ch(x)}{sh(x)}=\frac{e^x+e^{-x}}{e^x-e^{-x}}
\end{equation}
The hyperbolic cotangent function can be expressed with hyperbolic
cosecant function:
\begin{equation}
cth(x)=\sqrt{1+csch^2(x)}
\end{equation}
The inverse function of hyperbolic cosecant is denoted as
$Arcsch(x)$,and certainly $csch(Arcsch(x))=x$.
\section{Csch Cardassian for unitary matter and radiation}

We treat matter and radiation together for the  matter term, denoted
as $\rho_m$, and the total energy density of hyperbolic cosecant
(hereafter donated as Csch) Cardassian universe is, as above:
\begin{equation}
\rho'=\rho_m\left(1+b\cdot csch\left(\frac{\rho_m}{\rho_{card}}
\right)^n\right)
\end{equation}
with the Cardassian term:
\begin{equation}
f(\rho)=\rho_m b\cdot csch\left(\frac{\rho_m}{\rho_{card}} \right)^n
\end{equation}
$n$ and  $b$ are tuning parameters, which need observational
constraints. $\rho_{card}$ is a constant density value for
reference, as above in exponential Cardassian. when
$\rho_m\to\infty$, $\rho'\to\rho_m$, thus the modified Friedmann
equation returns to the normal form, and
\begin{equation}\label{eq:card4}
\frac{\partial f}{\partial \rho_m}=b\cdot
csch\left(\frac{\rho_m}{\rho_{card}} \right)^n-nb\cdot
csch\left(\frac{\rho_m}{\rho_{card}} \right)^n\cdot
cth\left(\frac{\rho_m}{\rho_{card}} \right)^n\cdot
\left(\frac{\rho_m}{\rho_{card}} \right)^n
\end{equation}
Hence from (\ref{eq:card3}) we get the pressure that:
$$
p=\rho_m\frac{\partial f}{\partial \rho_m}-f
$$
$$
=\rho_mb\cdot csch\left(\frac{\rho_m}{\rho_{card}}
\right)^n-n\rho_mb\cdot csch\left(\frac{\rho_m}{\rho_{crit}}
\right)^n\cdot cth\left(\frac{\rho_m}{\rho_{card}} \right)^n
\cdot\left(\frac{\rho_m}{\rho_{card}} \right)^n -\rho_mb\cdot
csch\left(\frac{\rho_m}{\rho_{card}} \right)^n
$$
\begin{equation}
=-bn\rho_m\cdot csch\left(\frac{\rho_m}{\rho_{crit}} \right)^n\cdot
cth\left(\frac{\rho_m}{\rho_{crit}} \right)^n
\cdot\left(\frac{\rho_m}{\rho_{crit}} \right)^n
\end{equation}
the Friedmann quation and Raychaudhuri equation in Csch cosmos is:
\begin{equation}
H^2=\frac{8\pi G}{3}\left[\rho_m+\rho_mb\cdot
csch\left(\frac{\rho_m}{\rho_{card}} \right)^n \right]
\end{equation}
and
$$
\dot{H}=-4\pi G\rho_m\left(1+\frac{\partial f}{\partial
\rho_m}\right)
$$
\begin{equation}
=-4\pi G\rho_m\left[1+b\cdot csch\left(\frac{\rho_m}{\rho_{card}}
\right)^n-nb\cdot csch\left(\frac{\rho_m}{\rho_{card}} \right)^n
cth\left(\frac{\rho_m}{\rho_{card}} \right)^n\cdot
\left(\frac{\rho_m}{\rho_{card}} \right)^n  \right]
\end{equation}
And the energy conservation equation $\dot{\rho'}=-3H(\rho'+p')$ is
equal to the conservation of particles' number
$\dot{\rho}_m=-3H\rho_m $ ,After the algebra above, we have known
the kinetic and dynamical equations of Csch Cardassian cosmos when
matter and radiation are treated together. Then we will research
into its phase space and autonomous equations. We define:
\begin{equation}
1=\frac{8\pi G}{3H^2}[\rho_m+f(\rho_m)]
\end{equation}
$$
\equiv \Omega_m+\Omega_{card}
$$
and introduce new variables $\kappa^2=8\pi G$:
\begin{equation}
x=\frac{\kappa\sqrt{\rho_m}}{\sqrt{3}H},\qquad
y=\frac{\kappa\sqrt{f}}{\sqrt{3}H}
\end{equation}
thus (\ref{eq:card4}) can be rewritten as:
\begin{equation}
\frac{\partial f}{\partial
\rho}=\frac{y^2}{x^2}-n\frac{y^2}{x^2}\cdot\sqrt{1+\left(\frac{y^2}{bx^2}\right)^2}\cdot
Arcsch\frac{y^2}{bx^2}
\end{equation}
According to (\ref{eq:card5}), we get the autonomous equations:
\begin{subequations}\label{eq:card6}
$$
\frac{dx}{dN}=-\frac{3}{2}x+\frac{3}{2}\left(1+\frac{\partial
f}{\partial \rho_m} \right)
$$
\begin{equation}
=-\frac{3}{2}x+\frac{3}{2}\left[\frac{y^2}{x^2}-n\frac{y^2}{x^2}\cdot\sqrt{1+\left(\frac{y^2}{bx^2}\right)^2}\cdot
Arcsch\frac{y^2}{bx^2}  \right]
\end{equation}
$$
\frac{dy}{dN}=-\frac{3}{2}\frac{x^2}{y}\frac{\partial f}{\partial
\rho_m}+\frac{3}{2}x^2y\left(1+\frac{\partial f}{\partial \rho_m}
\right)
$$
$$
=-\frac{3}{2}\left[\frac{y^2}{x^2}-n\frac{y^2}{x^2}\cdot\sqrt{1+\left(\frac{y^2}{bx^2}\right)^2}\cdot
Arcsch\frac{y^2}{bx^2}\right]
$$
\begin{equation}
+\frac{3}{2}x^2y\left[1+\frac{y^2}{x^2}-n\frac{y^2}{x^2}\cdot\sqrt{1+\left(\frac{y^2}{bx^2}\right)^2}\cdot
Arcsch\frac{y^2}{bx^2}\right]
\end{equation}
\end{subequations}
For flat FRW universe, $\Omega_m+\Omega_{card}=1$, and there's the
constraining equation $x^2+y^2=1$. So(\ref{eq:card6}) can be reduced
into:
$$
\frac{dx}{dN}=-\frac{3}{2}x+\frac{3}{2}\left[\frac{1}{x^2}-1-(\frac{1}{x^2}-1)\cdot\sqrt{1+\left(\frac{1-x^2}{bx^2}
\right)^2}\cdot Arcsch\frac{1-x^2}{bx^2} \right]
$$
\begin{equation}
=-\frac{3}{2}x^3-\frac{3}{2}x(1-x^2)\cdot\sqrt{1+\left(\frac{1-x^2}{bx^2}
\right)^2}\cdot Arcsch\frac{1-x^2}{bx^2}
\end{equation}
This is the very differential autonomous equation for Csch
Cardassian cosmos with unitary matter and radiation.

\section{Csch Cardassian for separate matter and radiation}
\subsection{Dynamical Equations}
In this section the energy density of matter and radiation will be
treat separately, so the total energy density is:
\begin{equation}
\rho'=(\rho_m+\rho_\gamma)\left[1+b\cdot
csch\left(\frac{\rho_m+\rho_\gamma}{\rho_{card}} \right)^n \right]
\end{equation}
thus the Csch Cardassian term is:
\begin{equation}
f(\rho_m,\rho_\gamma)=f(\rho_m+\rho_\gamma)=b(\rho_m+\rho_\gamma)\cdot
csch\left(\frac{\rho_m+\rho_\gamma}{\rho_{card}} \right)^n
\end{equation}
As for Csch function, the symmetric property $csch(\rho_m,
\rho_\gamma)=csch(\rho_m+\rho_\gamma)$ holds,so the matter term and
radiation term are symmetric mathematically to each other.
\begin{equation}\label{eq:card8}
\frac{\partial f}{\partial \rho_m}=\frac{\partial f}{\partial
\rho_\gamma}=b\cdot csch\left(\frac{\rho_m+\rho_\gamma}{\rho_{card}}
\right)^n-nb\cdot csch\left(\frac{\rho_m+\rho_\gamma}{\rho_{card}}
\right)^n\cdot cth\left(\frac{\rho_m+\rho_\gamma}{\rho_{card}}
\right)^n\cdot \left(\frac{\rho_m+\rho_\gamma}{\rho_{card}}
\right)^n
\end{equation}
According to \ref{eq:card7} the total pressure is:
$$
p'=\frac{\rho_\gamma}{3}+\rho_m\frac{\partial f}{\partial
\rho_m}+\frac{4}{3}\rho_\gamma\frac{\partial f}{\partial
\rho_\gamma}-f
$$
$$
=\frac{\rho_\gamma}{3}+(\rho_m+\frac{4}{3}\rho_\gamma)\frac{\partial
f}{\partial \rho_m}-f
$$
$$
=\frac{\rho_\gamma}{3}+(\rho_m+\frac{4}{3}\rho_\gamma) b\cdot
csch\left(\frac{\rho_m+\rho_\gamma}{\rho_{card}} \right)^n$$

$$-(\rho_m+\frac{4}{3}\rho_\gamma)nb\cdot
csch\left(\frac{\rho_m+\rho_\gamma}{\rho_{card}} \right)^n\cdot
cth\left(\frac{\rho_m+\rho_\gamma}{\rho_{card}} \right)^n\cdot
\left(\frac{\rho_m+\rho_\gamma}{\rho_{card}} \right)^n$$
 $$
-(\rho_m+\rho_\gamma)b(\rho_m+\rho_\gamma)\cdot
csch\left(\frac{\rho_m+\rho_\gamma}{\rho_{card}} \right)^n
$$
$$=\frac{\rho_\gamma}{3}+\frac{\rho_\gamma}{3}b\cdot
csch\left(\frac{\rho_m+\rho_\gamma}{\rho_{card}} \right)^n  $$
\begin{equation}
-(\rho_m+\frac{4}{3}\rho_\gamma)\cdot nb\cdot
csch\left(\frac{\rho_m+\rho_\gamma}{\rho_{card}} \right)^n\cdot
cth\left(\frac{\rho_m+\rho_\gamma}{\rho_{card}} \right)^n\cdot
\left(\frac{\rho_m+\rho_\gamma}{\rho_{card}} \right)^n
\end{equation}
With Friedmann equation and Raychaudhuri equation being:
\begin{equation}
H^2=\frac{8\pi G}{3}[\rho_m+\rho_\gamma+f(\rho_m,\rho_\gamma)]
\end{equation}
and
\begin{equation}
\dot{H}=-4\pi G(\rho_m+\frac{4}{3}\rho_\gamma+\rho_m\frac{\partial
f}{\partial \rho_m}+\frac{4}{3}\rho_\gamma\frac{\partial f}{\partial
\rho_\gamma})
\end{equation}
Here the energy conservation means the conservation of particles'
number and the radiation conservation:
\begin{equation}
\dot{\rho}_m=-3H\rho_m
\end{equation}
\begin{equation}
\dot{\rho}_\gamma=-4H\rho_\gamma
\end{equation}
Now we have got the dynamical equations for Csch Cardassian cosmos
with the radiation term and matter term treated separately. Then
we'll investigate the autonomous equations in its phase
space.
\subsection{Phase Space and Autonomous Equations}
From Friedmann equations, we define:
\begin{equation}
1=\frac{8\pi G}{3H^2}[\rho_m+\rho_\gamma+f(\rho_m,\rho_\gamma)]
\end{equation}
$$
\equiv \Omega_m+\Omega_\gamma+\Omega_{card}
$$
where $\Omega_m$,$\Omega_\gamma$,$\Omega_{card}$ is the relative
energy density for matter, radiation and Csch Cardassian term
respectively. Further more we introduce new variables with
$\kappa^2=8\pi G$:
\begin{equation}
x=\frac{\kappa \sqrt{\rho_m}}{\sqrt{3}H},\qquad y=\frac{\kappa
\sqrt{\rho_\gamma}}{\sqrt{3}H},\qquad z=\frac{\kappa
\sqrt{f}}{\sqrt{3}H},\qquad N=\ln a
\end{equation}
Thus (\ref{eq:card8}) reads:
\begin{equation}
\frac{\partial f}{\partial \rho_m}=\frac{\partial f}{\partial
\rho_\gamma}=\frac{z^2}{x^2+y^2}-n\cdot\frac{z^2}{x^2+y^2}\cdot\sqrt{1+\left(\frac{z^2}{b(x^2+y^2)}
 \right)^2}\cdot Arcsch\frac{z^2}{b(x^2+y^2)}
\end{equation}
For flat FRW cosmos with $\Omega_m+\Omega_\gamma+\Omega_{card}$, or
$x^2+y^2+z^2=1$, combined with the symmetry,
$f(\rho_m,\rho_\gamma)=f(\rho_m+\rho_\gamma)$,according to
(\ref{eq:card9}), we have the autonomous equations:
\begin{subequations}
\begin{equation}
\frac{dx}{dN}=-\frac{3}{2}x+\left(\frac{3}{2}x^3+2xy^2
\right)\left(1+\frac{z^2}{x^2+y^2}\right)
\end{equation}
$$
+\left(\frac{3}{2}x^3+2xy^2
\right)\left[-n\frac{z^2}{x^2+y^2}\sqrt{1+\left(\frac{1-x^2-y^2}{b(x^2+y^2)}
 \right)^2}\cdot Arcsch\frac{1-x^2-y^2}{b(x^2+y^2)}\right]
$$
\begin{equation}
\frac{dy}{dN}=-2y+\left(\frac{3}{2}x^2 y+2y^3 \right)
\frac{1-x^2-y^2}{x^2+y^2}
\end{equation}
$$
-\left(\frac{3}{2}x^2 y+2y^3
\right)\left[n\cdot\frac{1-x^2-y^2}{x^2+y^2}\cdot\sqrt{1+\left(\frac{1-x^2-y^2}{b(x^2+y^2)}
 \right)^2}\cdot Arcsch\frac{1-x^2-y^2}{b(x^2+y^2)} \right]
$$
\end{subequations}
\subsection{Critical Points}
From the above analysis, we get the autonomous equations in Csch
Cardassian for separate treatment of matter and energy density. The
following work is to get the  critical points. and we will show
there are three and only three critical points,
namely:$(0,0)(1,0)(0,1)$.\\

Obviously, $1>x>0$,$1>y>0$ and $1>x^2+y^2$ hold,firstly let's
introduce a new variable:
\begin{equation}
\phi=\frac{1-(x^2+y^2)}{x^2+y^2}>0
\end{equation}
and formulate a new function of:
\begin{equation}\label{eq:card-3}
\Psi=\phi-n\phi\sqrt{1+\phi^2/b^2}\cdot Arcsch\frac{\phi}{b}
\end{equation}
For (\ref{eq:card-3}), when $(x,y)\rightarrow (0,0)$,
$\phi\rightarrow +\infty$, $n\phi\sqrt{1+\phi^2/b^2}\rightarrow
\infty$, $Arcsch\frac{\phi}{b}\rightarrow 0$; and when
$(x,y)\rightarrow (1,0)$ or $(0,1)$, $\phi\rightarrow 0$,
$n\phi\sqrt{1+\phi^2/b^2}\rightarrow 0$,
$Arcsch\frac{\phi}{b}\rightarrow\infty$, The limit of $\Psi$ is
impossible to be read directly in both approximation cases. So we
transform $\Psi$ and employ the familiar Los'pital theorem (prime
operator ($'$)refers to derivative of the dependent variable to the
independent variable):
$$
\Psi=\phi-\frac{Arcsch\frac{\phi}{b}}{\frac{1}{n\phi\sqrt{1+\phi^2/b^2}}}
$$
$$
=\phi-\frac{\left(Arcsch\frac{\phi}{b}\right)'}{\left(\frac{1}{n\phi\sqrt{1+\phi^2/b^2}}\right)'}
$$
\begin{equation}\label{eq:card-4}
=\phi-\frac{\frac{n^2\phi^4}{b^2}+n^2\phi^2}{\frac{n\phi^3}{b^3}+\frac{n\phi^2}{b^3}+\frac{n\phi}{b}
}
\end{equation}
Based on this equation, let's consider two approximation cases:\\
 (1) When $(x,y)\rightarrow (0,0)$, $\phi\rightarrow\infty$:
\begin{equation}
\Psi=\phi-\phi\frac{\frac{n^2\phi^3}{b^2}+n^2\phi}{\frac{n\phi^3}{b^3}+\frac{n\phi^2}{b^3}+\frac{n\phi}{b}
}
\end{equation}
$$\sim\phi-nb\cdot\phi$$
So the infinitesimal $Arcsch\frac{\phi}{b}$ is magnified by the
infinity $n\phi\sqrt{1+\phi^2/b^2}$ to get an unitary infinity
$n\phi\sqrt{1+\phi^2/b^2}\cdot Arcsch\frac{\phi}{b}$, which is
coincidently of the same order with the infinity $\phi$, who will
conduct a minus operation with it. The minus of two infinity of the
same order can leads to conservance ! And we put $\phi$ back to the
original equation, and when  $(x,y)\rightarrow (0,0)$, we have:
\begin{equation}
\frac{dx}{dN}\sim\left[\frac{3}{2}x^3+2xy^2 \right]\cdot
(1-nb)\frac{1-(x^2+y^2)}{x^2+y^2}
\end{equation}
$$
>\left[\frac{3}{2}x^3+\frac{3}{2}xy^2 \right]\cdot
(1-nb)\frac{1-(x^2+y^2)}{x^2+y^2}
$$
$$
\sim\frac{3}{2}x\cdot (1-\frac{b^2}{2}n)[1-(x^2+y^2)]\quad\sim 0
$$
It's notable that  $\Psi$ is conserving itself if we put $1-nb=0$.
However, this constraint is too strong and we will have much less
room left for parameter tuning. Considering there is an
infinitesimal factor waiting to multiply $\Psi$, we'd better realize
the conservation in the next step, rather than set $1-nb=0$ in a
hurry in advance. Hence $(0,0)$ is a critical point of this autonomous system.\\

(2)When $(x,y)\rightarrow(1,0)$ or $(0,1)$, $\phi\rightarrow 0$, in
$$ \frac{\frac{n^2\phi^4}{b^2}+n^2\phi^2}{\frac{n\phi^3}{b^3}+\frac{n\phi^2}{b^3}+\frac{n\phi}{b}
}$$ the numerator is higher order infinitesimal compared to the
denominator, so the limit is zero, and:
$$\Psi=\phi-\frac{\frac{n^2\phi^4}{b^2}+n^2\phi^2}{\frac{n\phi^3}{b^3}+\frac{n\phi^2}{b^3}+\frac{n\phi}{b}
}$$ whose limit is also zero. Now let's go back to analyze the
original equation:
$$\frac{dy}{dN}\sim-2y+\left[\frac{3}{2}x^2y+2y^3
\right]\left[1+\phi-\frac{\frac{n^2\phi^4}{b^2}+n^2\phi^2}{\frac{n\phi^3}{b^3}+\frac{n\phi^2}{b^3}+\frac{n\phi}{b}
} \right] $$
\begin{equation}
\sim -2y+ \frac{3}{2}x^2 y+2y^3 \quad\sim 0
\end{equation}

Hence $(1,0)$,$(0,1)$ are both critical points of the autonomous system. \\

All in all, $(0,0)$, $(1,0)$, $(0,1)$ are three critical points of
this autonomous system. And because the critical points are
numerically achieved via solving the group of two three-order
algebraic equations, according to Gauss's basic theorem of
arithmetics, we could have only three roots in the complex domain.
So the three critical points achieved via approximation analyzing
are the only ones.
\subsection{Stability Matrix for Autonomous System}
Now we'll make phenomenological analysis of the stability of the
critical points. \\
Set $X=\frac{dx}{dN}$, $Y=\frac{dy}{dN}$, and the perturbation
matrix will be:
\begin{equation}
\mathcal{M}=\left(
\begin{array}{cc}
\frac{\partial X}{\partial x}\quad & \frac{\partial X}{\partial y}\\
\frac{\partial Y}{\partial x}\quad & \frac{\partial Y}{\partial y}
\end{array}
\right)
\end{equation}
The four elements of this Jacobian matrix is:
$$
\frac{\partial X}{\partial x}=-\frac{3}{2}+\left(\frac{9}{2}x^2+2y^2
\right)\left[1+\frac{1-(x^2+y^2)}{x^2+y^2} \right]
$$
$$
-\left(\frac{9}{2}x^2+2y^2
\right)\left[n\frac{1-(x^2+y^2)}{x^2+y^2}\sqrt{1+\left(\frac{1-(x^2+y^2)}{b(x^2+y^2)}
\right)^2}Arcsch\frac{1-(x^2+y^2)}{b(x^2+y^2)}\right]
$$
$$
+\left(\frac{3}{2}x^3+2xy^2\right)
n\frac{2x}{(x^2+y^2)^2}\sqrt{1+\left(\frac{1-(x^2+y^2)}{b(x^2+y^2)}
 \right)^2}Arcsch\frac{1-(x^2+y^2)}{b(x^2+y^2)}
$$
$$
+n\left(\frac{3}{2}x^3+2xy^2\right)\frac{(1-(x^2+y^2))^2}{(x^2+y^2)^4}\frac{2x}{b^2\sqrt{1+\frac{1}{b^2}(\frac{1}{x^2+y^2}-1)^2}}Arcsch\frac{1-(x^2+y^2)}{b(x^2+y^2)}
$$
\begin{equation}
-\frac{2x}{(x^2+y^2)^2}\cdot\left(\frac{3}{2}x^3+2xy^2\right)+\left(\frac{3}{2}x^3+2xy^2\right)n\frac{2x}{(x^2+y^2)^2}
\end{equation}
and$$ \frac{\partial X}{\partial
y}=4xy\left[1+\frac{1-(x^2+y^2)}{x^2+y^2}\right]
$$
$$
-4xy\cdot
n\frac{1-(x^2+y^2)}{x^2+y^2}\sqrt{1+\left(\frac{1-(x^2+y^2)}{b(x^2+y^2)}
\right)^2}Arcsch\frac{1-(x^2+y^2)}{b(x^2+y^2)}
$$
$$
+\left(\frac{3}{2}x^3+2xy^2\right)
n\frac{2y}{(x^2+y^2)^2}\sqrt{1+\left(\frac{1-(x^2+y^2)}{b(x^2+y^2)}
 \right)^2}Arcsch\frac{1-(x^2+y^2)}{b(x^2+y^2)}
$$

$$
+\left(\frac{3}{2}x^3+2xy^2\right)n\frac{(1-(x^2+y^2))^2}{(x^2+y^2)^4}\frac{2y}{b^2\sqrt{1+\frac{1}{b^2}(\frac{1}{x^2+y^2}-1)^2}}Arcsch\frac{1-(x^2+y^2)}{b(x^2+y^2)}
$$
\begin{equation}
-\frac{2y}{(x^2+y^2)^2}\left(\frac{3}{2}x^3+2xy^2\right)+\left(\frac{3}{2}x^3+2xy^2\right)n\frac{2y}{(x^2+y^2)^2}
\end{equation}

and$$ \frac{\partial Y}{\partial
x}=3xy\left[1+\frac{1-(x^2+y^2)}{x^2+y^2}\right]
$$
$$
-3xy\cdot
n\frac{1-(x^2+y^2)}{x^2+y^2}\sqrt{1+\left(\frac{1-(x^2+y^2)}{b(x^2+y^2)}
\right)^2}Arcsch\frac{1-(x^2+y^2)}{b(x^2+y^2)}
$$

$$+\left(\frac{3}{2}x^2y+2y^3\right)
n\frac{2x}{(x^2+y^2)^2}\sqrt{1+\left(\frac{1-(x^2+y^2)}{b(x^2+y^2)}
 \right)^2}Arcsch\frac{1-(x^2+y^2)}{b(x^2+y^2)}
$$

$$
+\left(\frac{3}{2}x^2y+2y^3\right)n\frac{(1-(x^2+y^2))^2}{(x^2+y^2)^4}\frac{2x}{b^2\sqrt{1+\frac{1}{b^2}(\frac{1}{x^2+y^2}-1)^2}}Arcsch\frac{1-(x^2+y^2)}{b(x^2+y^2)}
$$
\begin{equation}
-\frac{2x}{(x^2+y^2)^2}\left(\frac{3}{2}x^2y+2y^3\right)+n\left(\frac{3}{2}x^2y+2y^3\right)\frac{2x}{(x^2+y^2)^2}
\end{equation}

and$$ \frac{\partial Y}{\partial y}=-2+\left(-\frac{3}{2}x^2+6y^2
\right)\left[1+\frac{1-(x^2+y^2)}{x^2+y^2}\right]$$

$$-\left(-\frac{3}{2}x^2+6y^2
\right)n\frac{1-(x^2+y^2)}{x^2+y^2}\sqrt{1+\left(\frac{1-(x^2+y^2)}{b(x^2+y^2)}
\right)^2}Arcsch\frac{1-(x^2+y^2)}{b(x^2+y^2)}$$

$$+\left(\frac{3}{2}x^2y+2y^3\right)
n\frac{2y}{(x^2+y^2)^2}\sqrt{1+\left(\frac{1-(x^2+y^2)}{b(x^2+y^2)}
 \right)^2}Arcsch\frac{1-(x^2+y^2)}{b(x^2+y^2)}$$

$$
+\left(\frac{3}{2}x^2y+2y^3\right)n\frac{(1-(x^2+y^2))^2}{(x^2+y^2)^4}\frac{2x}{b^2\sqrt{1+\frac{1}{b^2}(\frac{1}{x^2+y^2}-1)^2}}Arcsch\frac{1-(x^2+y^2)}{b(x^2+y^2)}
$$

\begin{equation}
-\frac{2y}{(x^2+y^2)^2}\left(\frac{3}{2}x^2y+2y^3\right)+\left(\frac{3}{2}x^2y+2y^3\right)n\frac{2y}{(x^2+y^2)^2}
\end{equation}

\chapter{Virial Collapse of Power-style Cardassian}
\section{Applications of Virialization in Cosmology}
The spherical collapse mechanism, which was put forward by Gunn and
Gott \cite{GunnGot} \cite{GunnGot2}, is an effective tool in
understanding the consequence of the density inhomogeneity. An
overdensed volume will slow down from the co-expansion, until the
expanding velocity vanishes. Then it starts to collapse over
rotating, and potential energy translates into rotating kinetic
energy. The special state when the expanding just stops and collapse
is to start is dubbed turn around state. Ultimately the collapse
ends up with stable virial structure.\\

We can apply the virial theorem to the above process in the light of
energy conservation. A general statement of virial theorem says
that, during a long period of time, the opposite value of the
average kinetic energy of a particle group equals to the virial
force acting on it:
\begin{equation}
\bar{T}=-\frac{1}{2}\sum_{i}\vec{F}_i\cdot\vec{r}_i
\end{equation}
where $\bar{T}$ is the total rotating kinetic energy, $\vec{F}_i$
and $\vec{r}_i$ refer to the external forces and location vector of
the $i$th particle. For conserving system,
\begin{equation}
\bar{T}=\frac{1}{2}\bar{\sum_i(\nabla_iV)\cdot \vec{r}_i}
\end{equation}
where $V$ is the potential of the particle group, $\nabla$ is the
vectoring gradient operator:
\begin{equation}
\nabla_i=\frac{\partial}{\partial
x_i}\hat{i}+\frac{\partial}{\partial
y_i}\hat{j}+\frac{\partial}{\partial z_i}\hat{k}
\end{equation}

The total potential energy, the total virial potential energy and
the static radius of the virialized state at the turn around (when
the expansion stops, and  collapse and rotating takes place) are
denoted as $U_{ta}$, $U_{vir}$ and $R_{vir}$ respectively. Supposing
the system's potential energy takes the analytic form of $U$, then
the kinetic energy after virialization, which depends on $U$, is:
\begin{equation}
T_{vir}=\big(\frac{R}{2}\frac{\partial U}{\partial R}\big)_{vir}
\end{equation}
The energy conserves, so after the virial collapse(T=0 when collapse
starts):
\begin{equation}
\left[U+\frac{R}{2}\frac{\partial U}{\partial R}\right]_{vir}=U_{ta}
\end{equation}
The analytic expression of the potential energy term U can be
deduced as follows: Supposing $\rho_Q$ is the energy density for the
virial component, $\Phi_Q$ is the total potential acts on a certain
point in the sphere, and the total potential energy will be
\begin{equation}
U=\frac{1}{2}\int\rho_Q\Phi_Q dV
\end{equation}
$$=\frac{1}{2}\int\rho_Q\Phi_Q(r)d(\frac{4}{3}\pi r^3)  $$
$$=\frac{1}{2}\int\rho_Q\Phi_Q(r)\cdot 4\pi r^2 dr$$
In a homogeneous sphere, $\rho_Q$ satisfies:
\begin{equation}
\dot{\rho}_Q+3(1+w_Q)\frac{\dot{r}}{r}\rho_Q=0
\end{equation}
And the total potential in the sphere for a certain point is:
\begin{equation}
\Phi_Q(r)=-2\pi G(1+3w_Q)\rho_Q\left(R^2-\frac{r^2}{3}\right)
\end{equation}
Where $R$ is the radius of the sphere, $r$ is the distance between a
point and the spherical center.
\section{Virialization Power-style Cardassian}
Now let's consider the virialization of power-style
Cardassian\cite{Card1}. Supposing that Cardassian term and matter
term both participate the virialization process. i.e.
\begin{equation}
\rho=\rho_m+b\rho_m^n,\qquad n<2/3
\end{equation}
\subsection{Virialization of Matter Term}
EOS parameter for matter term is $w=0$, and the distribution of
potential in the sphere is:
\begin{equation}
\Phi(r)=-2\pi G(1+3\times 0)\rho_m\left(R^2-\frac{r^2}{3}\right)
\end{equation}
$$=-2\pi G\rho_m\left(R^2-\frac{r^2}{3}\right)$$
The total potential energy in the sphere of radius $R$ is:
\begin{equation}
U_m=-\int^R_0\rho^2_m\cdot 2\pi G\cdot 2\pi
r^2\cdot\left(R^2-\frac{r^2}{3}\right)dr
\end{equation}
$$=-\frac{16}{15}\pi^2G\rho_m^2R^5$$
The rotating kinetic energy associating $U_m$ is:
\begin{equation}\label{eq:virial-1}
T_m=\frac{R}{2}\frac{\partial U_m}{\partial
R}=-\frac{8}{2}\pi^2G\rho_m^2R^5
\end{equation}
Obviously $T_m<0$ in (\ref{eq:virial-1}). Kinetic energy cannot be
negative, so the virial collapse of matter term alone is forbidden.
\subsection{Virialization GF Fluid/Cardassian Term}
Now let's have a look at the behaviors of Cardassian energy.
Power-style Cardassian is perfect fluid with EOS parameter $w=n-1$
and the distribution of potential in a virializing sphere is:
\begin{equation}
\Phi(r)=-2\pi G(3n-2)(b\rho_m^n)\left(R^2-\frac{r^2}{3}\right)
\end{equation}
So the total potential energy in the sphere is:
\begin{equation}
U=\frac{1}{2}\int\rho\Phi(r)\cdot 4\pi r^2 dr
\end{equation}
$$=-\frac{1}{2}\int_0^R\rho_m\cdot 2\pi
G(3n-2)(b\rho_m^n)\left(R^2-\frac{r^2}{3}\right)\cdot 4\pi r^2 dr
$$
$$=\frac{16}{15}G\pi^2 R^5(2-3n)b\rho_m^{2n}  $$
and the associated virial rotating energy is:
\begin{equation}
T_{vir}=\frac{R_{vir}}{2}\frac{\partial U}{\partial
R}=\frac{8}{3}G\pi^2
R_{vir}^5(2-3n)(\rho_{m,vir}+b\rho_{m,vir}^n)^2=\frac{5}{2}U_{vir}
\end{equation}
Energy is conserving during the virialization process.
\begin{equation}
\frac{7}{2}\times\frac{16}{15}G\pi^2
R_{vir}^5(2-3n)(b\rho_{m,vir}^{2n})=\frac{16}{15}G\pi^2
R_{ta}^5(2-3n)(b\rho_{m,ta}^{2n})
\end{equation}
reduced to be:
\begin{equation}\label{eq:card-energy-virial-conserve}
\frac{7}{2}R_{vir}^5\rho_{m,vir}^{2n}=R_{ta}^5\rho_{m,ta}^{2n}
\end{equation}
The mass does't lose, so
\begin{equation}\label{eq:card-mass-virial-conserve}
\frac{\rho_{m,vir}}{\rho_{m,ta}}=\left(\frac{R_{ta}}{R_{vir}}\right)^3
\end{equation}
Put \ref{eq:card-mass-virial-conserve}
into\ref{eq:card-energy-virial-conserve} to get:
\begin{equation}
\left(\frac{R_{vir}}{R_{ta}}
\right)^5\left(\frac{\rho_{m,vir}}{\rho_{m,ta}}
\right)^{2n}=\left(\frac{R_{vir}}{R_{ta}}
\right)^5\left((\frac{R_{ta}}{R_{vir}})^3
\right)^{2n}=\left(\frac{R_{vir}}{R_{ta}} \right)^{5-6n}=\frac{2}{7}
\end{equation}
Hence the ratio of the stable virial radius with turn around radius
is:
\begin{equation}
\frac{R_{vir}}{R_{ta}}=\sqrt[5-6n]{\frac{2}{7}}
\end{equation}
The upper calculations constrains parameter $n$ for twice, namely
kinetic energy being positive and collapse radius becoming smaller
than initial:
\begin{subequations}
\begin{equation}
T>0\quad\sim 2-3n>0
\end{equation}
\begin{equation}
\frac{R_{vir}}{R_{ta}}=\sqrt[5-6n]{\frac{2}{7}}<1
\end{equation}
\end{subequations}
So, after virialization GF fluids collapse to a steady sphere with
certain radius which depends on the initial radius and the
Cardassian parameter $n$. And no singularity is generated.
\subsection{Unitary Virialization of Matter Term and GF Fluid Term}
Although matter term $\rho_m$ cannot virialize on itself, howerever,
as it is shown below, due to the excellent virialization property of
GF fluid, the matter term $\rho_m$ can virialize together with GF
fluid $\rho_{card}$. This is pretty similar to the fact that high
energy $\gamma$ photon cannot decay into electron-positron pairs
when isolated,
but will decay if located in Column field nearby the nucleus. \\

As analyzed in chapter 6, that the total pressure of $\rho_m$ and
$\rho_{card}$ is $p=(n-1)b\rho_m^n$, so the EOS parameter is:
\begin{equation}
w=\frac{(n-1)b\rho_m^n}{\rho_m+b\rho_m^n}
\end{equation}
The distribution of potential generated by $\rho_m$ and
$\rho_{card}$ in the virial sphere is:
\begin{equation}
\Phi(r)=-2\pi G\left(1+\frac{3(n-1)b\rho_m^n}{\rho^n_m} \right)
(\rho_m+b\rho_m^n )\left(R^2-\frac{r^2}{3} \right)
\end{equation}
$$-2\pi G\big(\rho_m+3(n-2)b\rho_m^n\big)$$
$\rho_m$ and $\rho_{card}$ leads to the potential:
\begin{equation}
U=\frac{1}{2}\int^r_m(-2\pi G)\big(\rho_m+3(n-2)b\rho_m^n\big)dr
\end{equation}
$$=\frac{16}{15}\pi^2GR^5(\rho_m+b\rho_m^n)\big((2-3n)b\rho_m^n-\rho_m\big) $$
the associating rotating energy:
\begin{equation}
T=\frac{8}{3}\pi^2GR^5(\rho_m+b\rho_m^n)\big((2-3n)b\rho_m^n-\rho_m\big)=\frac{5}{2}U
\end{equation}
Thus, $T>0$ is possible if $b,n$ are properly tuned. Due to energy
conserving:
\begin{equation}
\frac{7}{2}\times\frac{16}{15}\pi^2GR^5_{vir}(\rho_{m,vir}+b\rho_{m,vir}^n)\big((2-3n)b\rho_m^n-\rho_m\big)
\end{equation}
$$=\frac{16}{15}\pi^2GR^5_{ta}(\rho_{m,ta}+b\rho_{m,ta}^n)\big((2-3n)b\rho_m^n-\rho_m\big)$$
The mass is conserving
\begin{equation}
\frac{\rho_{m,vir}}{\rho_{m,ta}}=\left(\frac{R_{ta}}{R_{vir}}\right)^3
\end{equation}
Combine it with energy conserving equation:
\begin{equation}
\frac{7}{2}\left(\frac{R_{vir}}{R_{ta}}
\right)^5=\frac{(\rho_{m,ta}+b\rho_{m,ta}^n)\big((2-3n)b\rho_m^n-\rho_m\big)}{(\rho_{m,vir}+b\rho_{m,vir}^n)\big((2-3n)b\rho_m^n-\rho_m\big)}
\end{equation}
$$=\frac{(2-3n)b^2\rho^{2n}_{m,ta}+(1-3n)b\rho^{n+1}_{m,ta}-\rho_{m,ta}^2}{(2-3n)b^2\rho^{2n}_{m,vir}+(1-3n)b\rho^{n+1}_{m,vir}-\rho_{m,vir}^2}$$
Set the ratio $\frac{R_{vir}}{R_{ta}}=x$, and via the mass
conservation,
\begin{equation}\label{eq:virial-4}
\frac{7}{2}x^5=\frac{(2-3n)b^2\rho^{2n}_{m,vir}x^{6n}+(1-3n)b\rho^{n+1}_{m,vir}x^{3n+3}-\rho_{m,vir}^2x^6}{(2-3n)b^2\rho^{2n}_{m,vir}+(1-3n)b\rho^{n+1}_{m,vir}-\rho_{m,vir}^2}
\end{equation}
Make deduction of $x^5$ to (\ref{eq:virial-4}) to get
$$\frac{7}{2}\left\{(2-3n)b^2\rho^{2n}_{m,vir}+(1-3n)b\rho^{n+1}_{m,vir}-\rho_{m,vir}^2\right\}$$
\begin{equation}\label{eq:virial-2}
=(2-3n)b^2\rho^{2n}_{m,vir}x^{6n-5}+(1-3n)b\rho^{n+1}_{m,vir}x^{3n-2}-\rho_{m,vir}^2x
\end{equation}
Introduce a new variable $q\equiv
\frac{\rho_{card,vir}}{\rho_{m,vir}}=\frac{b\rho_{m,vir}^n}{\rho_{m,vir}}$,
and
\begin{subequations}\label{eq:virial-3}
\begin{equation}
b^2\rho^{2n}_{m,vir}=q^2\rho^{2}_{m,vir}
\end{equation}
\begin{equation}
b^2\rho^{n+1}_{m,vir}=q\rho^{2}_{m,vir}
\end{equation}
\end{subequations}
Insert (\ref{eq:virial-3}) into (\ref{eq:virial-2}) to get
$$\frac{7}{2}\left\{(2-3n)q^2\rho^{2}_{m,vir}+(1-3n)q\rho_{m,vir}-\rho_{m,vir}^2\right\}$$
\begin{equation}\label{eq:virial-5}
=(2-3n)q^2\rho^{2}_{m,vir}x^{6n-5}+(1-3n)q\rho^{2}_{m,vir}x^{3n-2}-\rho_{m,vir}^2x
\end{equation}
Obviously $\rho_{m,vir}^2>0$ holds, and deduct it from both sides of
the equation (\ref{eq:virial-5}), and we get the relation of $x$ and
$q$:
$$\frac{7}{2}\left\{(2-3n)q^2 +(1-3n)q-1\right\}$$
\begin{equation}\label{eq:virial-6}
=(2-3n)q^2x^{6n-5}+(1-3n)qx^{3n-2}-x
\end{equation}
We can also get the relation $z_{vir}$ and $x$ via
(\ref{eq:virial-6}). Currently, Friedmann equation reads:
\begin{equation}
H^2_0=\frac{8\pi G}{3}(\rho_{m,0}+b\rho_{m,0}^2)
\end{equation}
i.e
\begin{equation}
\rho_{crit,0}\equiv \frac{3H_0^2}{8\pi G}=\rho_{m,0}+b\rho_{m,0}^2
\end{equation}
i.e
\begin{equation}
\frac{\rho_{crit,0}}{\rho_{m,0}}=1+\frac{b\rho_{m,0}^n}{\rho_{m,0}}
\end{equation}
i.e
\begin{equation}
\frac{1}{\Omega_{m,0}}-1=q
\end{equation}
insert it into $x-q$ relation (\ref{eq:virial-6}) and one will get
the $q-z_{vir}$ relation.


\section{Conclusions and Evaluation}
New work gathers in Chapter 6 and Chapter 7. Firstly we make a
comprehensive analysis of the existing work, pay particular
attention to Gondolo and Freese's idea of treating Cardassian energy
term as relativistic perfect fluid(GF fluid), point out that a
potential Cardassian term should meet three conditions, and review
three existing Cardassian terms (power style, polytropic style and
its modification, exponential style and its modification), and
eventually put forward the newly found hyperbolic cosecant
Cardassian.\\

Then the Cardassian dynamical equations are introduced generally and
logically under GF fluid scenario, together with the flowing process
of constructing phase space and differential dynamical systems from
Friedmann equation. Hyperbolic cosecant Cardassian term is employed
for concrete computing. The analysis proceeds in two cases, namely a
unified description of matter and radiation energy density (case 1)
and a separate description of matter and radiation terms (case 2).
Formalism of case 2 is more exact at the expense of more
complicatedness, and due to the mathematical symmetry of matter term
and radiation term in hyperbolic cosecant function, the differential
dynamical equations are considerably simplified. Phase space and
dynamical systems for both cases are achieved. When we calculate the
critical points for case 2, amazingly interesting behaviors of
self-consistency and auto-normalization are exhibited, which is a
strong support for the new model,
along with a forever positive sound speed.\\

The process of virial collapse in Cardassian cosmos is analyzed.
Power-style Cardassian term is employed for its simplicity.
Calculation declares that virial collapse of matter alone is
forbidden. Yet Cardassian has excellent ability for virial collapse,
after the virial collapse ending up with a stable sphere, the ratio
of the ultimate radius to the original radius depends on the
adjustable parameters in Cardassian term. And, the mixture of GF
fluid and matter could conduct virial collapse, the ratio of the
ultimate radius to the original radius depends on the adjustable
parameters in Cardassian term, too.

The creative work in this thesis incudes the introduction of
hyperbolic cosecant Cardassian, which is the fourth ever available
Cardassian term ever found, after the introduction of exponential
style in 2005. This work, along with the calculating of power
Cardassian virialization, enriches the research of Cardassian
cosmology.\\


\end{document}